\documentclass[sn-mathphys-ay]{sn-jnl}
\usepackage{amsthm,amsmath,amsfonts,amssymb}

\usepackage{mathtools,breqn,comment,graphicx}
\usepackage{xcolor,epstopdf}
\usepackage[title]{appendix}

\usepackage{amsthm,bm}
\newtheorem{property}{Property} 

\usepackage{amsmath,amsthm,mathrsfs,amsfonts}
\usepackage{enumerate}
\usepackage{natbib}
\usepackage{nicefrac}

\def\rank{\mathrm{rank}}
\def\ng{gamma--normal}
\def\en{exponential--normal}
\def\ngsymbol{\mathrm{GN}}
\def\ensymbol{\mathrm{EN}}
\def\odchi{overdispersed chi--squared}
\def\odchisymbol{\mathrm{B}}

\newcommand\Var{\text{Var}}

\newcommand\E{\text{E}}

\def\erf{\text{erf}}
\def\python{\tt{python}\rm}
\def\github{\tt{GitHub}\rm}
\def\ml{maximum likelihood}

\def\DP{$D_P$}

\def\dzdDi{\left[ \dfrac{\partial}{\partial \zeta} \left(\dfrac{D_{-r}'(\zeta_i)}{D_{-r}(\zeta_i)}\right) \right]}
\def\dzdsigma{\left( \alpha - \dfrac{\mu - z}{\sigma^2} \right)}
\def\DpD{\left(\dfrac{D_{-r}'(\zeta)}{D_{-r}(\zeta)}\right)} 
\def\DpDi{\left(\dfrac{D_{-r}'(\zeta_i)}{D_{-r}(\zeta_i)}\right)}
\def\ArA{\dfrac{A_r(\zeta,r)}{A(\zeta,r)}}
\def\ArAi{\left( \dfrac{A_r(\zeta_i,r)}{A(\zeta_i,r)}\right)}
\begin{document}

\title[Gamma--normal distribution]{Properties and \ml\ estimation of the \ng\ and related probability distributions}

\author*[1]{\fnm{Massimiliano} \sur{Bonamente}}\email{bonamem@uah.edu}
\author[2]{\fnm{Dale} \sur{Zimmerman}}\email{dale-zimmerman@uiowa.edu}
\affil*[1]{\orgdiv{Department of Physics and Astronomy}, \orgname{University of Alabama in Huntsville},\orgaddress{\street{301 Sparkman Dr.}, \city{Huntsville}, \postcode{35899}, \state{Alabama}, \country{U.S.A.}}}
\affil[2]{\orgdiv{Department of Statistics and Actuarial Science}, \orgname{University of Iowa}, 
\orgaddress{\city{Iowa City}, \postcode{52242}, \state{Iowa}, \country{U.S.A.}}}







\abstract{
This paper presents likelihood--based inference methods for the family of univariate
\ng\ distributions $\ngsymbol(\alpha, r, \mu, \sigma^2)$
  that result
  from summing independent $\gamma(\alpha, r)$ and $N(\mu,\sigma^2)$ random variables. 
  First, the probability density function of a \ng\ variable
  is provided in compact form with the use of parabolic cylinder functions,
  along with
  key properties. We then provide analytic expressions for the maximum--likelihood score equations and the Fisher information matrix, and discuss
  inferential methods for the \ng\ distribution.
  Given the widespread use of the two constituting distributions,
  the \ng\ distribution is a general purpose tool for a variety of applications.
  In particular, we discuss two distributions  that are obtained as special cases
  and that are featured in a variety of statistical applications:
  the \en\ distribution 
  and the chi--squared--normal (or \odchi) distribution. 
}


\maketitle

\section{Introduction}
\label{sec:intro}

The sum of two random variables occurs frequently in statistical applications. 
For example, it is
often required  to modify certain distributions with the addition of an independent normal distribution,
as a simple means to shift the mean and/or increase the variance of the original distribution, or to model
a given signal as the sum of two independent components, such as the source and the background.

For this class of applications, the exponential--normal (or exponential--Gaussian) distribution, 
which is the distribution of $Z=X+Y$ when
$X$ is an exponential random variable and $Y$ an independent normal random variable, has been used in
a variety of disciplines including
chromatography \citep{delley1985,gruskha1972}, cellular biology \citep[e.g.][]{golubev2010},
finance \citep[e.g.][]{carr2009} and psychology \citep[e.g.][]{palmer2011}.
The \ng\ distribution was also proposed as a generalization of the exponential--normal
distribution by \cite{plancade2012}, for the specific task of proper
background subtraction in certain biological applications \citep[e.g.][]{xie2009, wang2012}.
Moreover, the distribution of the sum of independent normal and chi--squared variables also occurs as the parent distribution
for the goodness--of--fit
statistic in Poisson regression with systematic errors, as was previously shown 
by
\cite{bonamente2023}, i.e., the \odchi\ distribution.

While the convolution of virtually any two distributions can be carried out numerically, there
are advantages to having an analytic form for the probability density function (pdf)
of the sum of two random variables. One of these is  
computational speed and precision. In fact, the computational 
cost of the 
convolution of two distributions 
is typically $O(n^2)$, where $n$ is the number of
samples in the convolution \citep[e.g.][]{karas2013}, and this may become prohibitively high in
certain applications that require high precision.
Second, an explicit compact form for the density makes it easier
to identify the role played by the parameters, 
thus making the distribution easier to use, interpret, and estimate.

Accordingly, the goal of this paper is two--fold. First, it presents a compact form for
the pdf and  properties of the family of \ng\ $\ngsymbol(\alpha,r,\mu,\sigma^2)$
random variables that result from the sum of a gamma $\gamma(\alpha, r)$ random variable and
an independent normal $N(\mu, \sigma^2)$ random variable.~\footnote{See App.~\ref{sec:appA}
for parameterization of the distributions and other mathematical properties.} Although the problem is elementary in its
methods, a compact form for the convolution of these two distributions
has not previously reported in the literature \citep[see, e.g., the comment after Eq.~6 of][]{plancade2012}. {This paper provides, for the first time, a closed form for the distribution
of the univariate \ng\ variable as a function of parabolic cylinder functions and other elementary functions.}
Given the wide use of the two constituting distributions, the
univariate normal--gamma distribution is therefore a convenient general--purpose statistical tool{, and the closed form provided in this paper is aimed to further its use}.

Secondly, the paper provides \ml\ score functions and the information matrix 
in analytic forms that are suitable for parameter {and error} estimation, and discusses  possible
applications for this family of distributions, including the special cases of the
\en\ $\ensymbol(\alpha, \mu,\sigma^2)$ and the
overdispersed $\chi^2$ distribution $\odchisymbol(r, \mu, \sigma^2)$. {
\cite{plancade2012} considered only a special case of the \ml\ estimation with the \ng\ distribution (i.e.,
as an application to data with separate source and background measurements), and did not develop
analytic expression for the score equations and the information matrix. In this paper we extend those
results to the general case. In particular, the availability of the information matrix makes it possible
to obtain estimates of the covariance matrix, which were not available in the \cite{plancade2012} method.}

This paper is structured as follows. Section~\ref{sec:ng} presents the probability
distribution function and key
properties of the \ng\ distribution. Section~\ref{sec:special} discusses two
special cases, viz. the \en\ and the \odchi\ distributions. Section~\ref{sec:ML} then
describes likelihood-based methods for the estimation of the \ng\ and associated distributions.
A brief review of applications of these distributions is then provided in
Section~\ref{sec:application}, followed by our conclusions in 
Section~\ref{sec:discussion}.

\section{The univariate \ng\ random variable}
\label{sec:ng}

We start by defining the normal--gamma variable as the sum of independent
gamma and normal random variables (one of each). A similar name (normal--gamma or gamma--normal) 
is often used to describe a
bivariate distribution with pdf 
given by the product of the gamma and normal pdfs,
which is of common use in Bayesian statistics \citep[e.g., ][p. 136]{bernardo2000}. To distinguish
between the two different distributions, the family of distributions under investigation in this
paper is referred to as the \textit{univariate} \ng\ distribution.
The qualifier will often be omitted for the sake of brevity, since all distributions
discussed in this paper are univariate.

\subsection{Definition}
Let $X\sim \gamma(\alpha, r)$ and $Y\sim N(\mu, \sigma^2)$ be two independent random variables, having
respectively a gamma distribution with rate parameter $\alpha$ and shape 
parameter $r$, and a normal distribution with mean $\mu$ and variance $\sigma^2$. 
Let $Z=X+Y$ be the sum of the
 two variables. The
 pdf of $Z$ is given by the convolution of the two densities,
 \begin{equation}
 \begin{aligned}
     f_{Z}(z; \alpha, r, \mu, \sigma^2) = & f_{\gamma} (z; \alpha, r) *  f_N(z; \mu, \sigma^2) \\
        = &  \int_{-\infty}^{\infty} f_{\gamma}(z-y; \alpha, r) f_N(y; \mu, \sigma^2)\, dy, \; \text{ for $z-y>0$}. 
 \end{aligned}
  \label{eq:conv}
\end{equation}
With the substitution $x=z-y>0$, the pdf of $Z$ becomes
\begin{equation}
  f_{Z}(z; \alpha, r, \mu, \sigma^2) = \dfrac{\alpha^r}{\Gamma(r)} \dfrac{1}{\sqrt{2 \pi \sigma^2}} \int_0^{\infty}  
  x^{r-1} e^{-\alpha x} e^{-{(x-(z-\mu))^2}/{2 \sigma^2}}\, dx, \; \text{ for $z \in \mathbb{R}$.}
  \label{eq:fA2}
\end{equation}

The variable $Z$ with pdf given by \eqref{eq:fA2} is said to belong to the family of \textit{univariate \ng} random variables, and it will
be indicated as
\begin{equation}
    Z \sim \ngsymbol(\alpha, r, \mu,\sigma^2).
\end{equation}
$Z$ is a real--valued variable, $z \in \mathbb{R}$, with parameters $\alpha, r \in \mathbb{R}^+$, $\mu \in  \mathbb{R}$,
and $\sigma^2 \in \mathbb{R}^+$.

\subsection{The probability density function}
The pdf of a $Z \sim \ngsymbol(\alpha, r, \mu,\sigma^2)$ \ng\ variable can be written as
a function of {parabolic cylinder functions} $D_p(z)$, which 
are solutions of the Weber differential equation \citep{weber1869} that results from 
separating the variables of the wave equation in parabolic cylindrical coordinates \citep[see also][]{Whittaker1902}.
The parabolic cylinder functions can be written in integral form as
\begin{equation}
  D_p(z) = \dfrac{e^{-z^2/4}}{\Gamma(-p)} \int_0^{\infty} e^{-xz -x^2/2} x^{-1-p} dx, \text{ for real $p<0$}.
  \label{eq:PCF}
\end{equation}
In general $z$ may be a complex number in the argument of the parabolic cylinder function, 
but for this application it is only interesting to consider real values of the variable.
An alternative notation for the
parabolic cylinder functions is
$U(a,z)=D_{-a-\nicefrac{1}{2}}(z)$ \citep[e.g.][]{Miller1952}.

\begin{figure}
    \centering
    \includegraphics[width=4.2in]{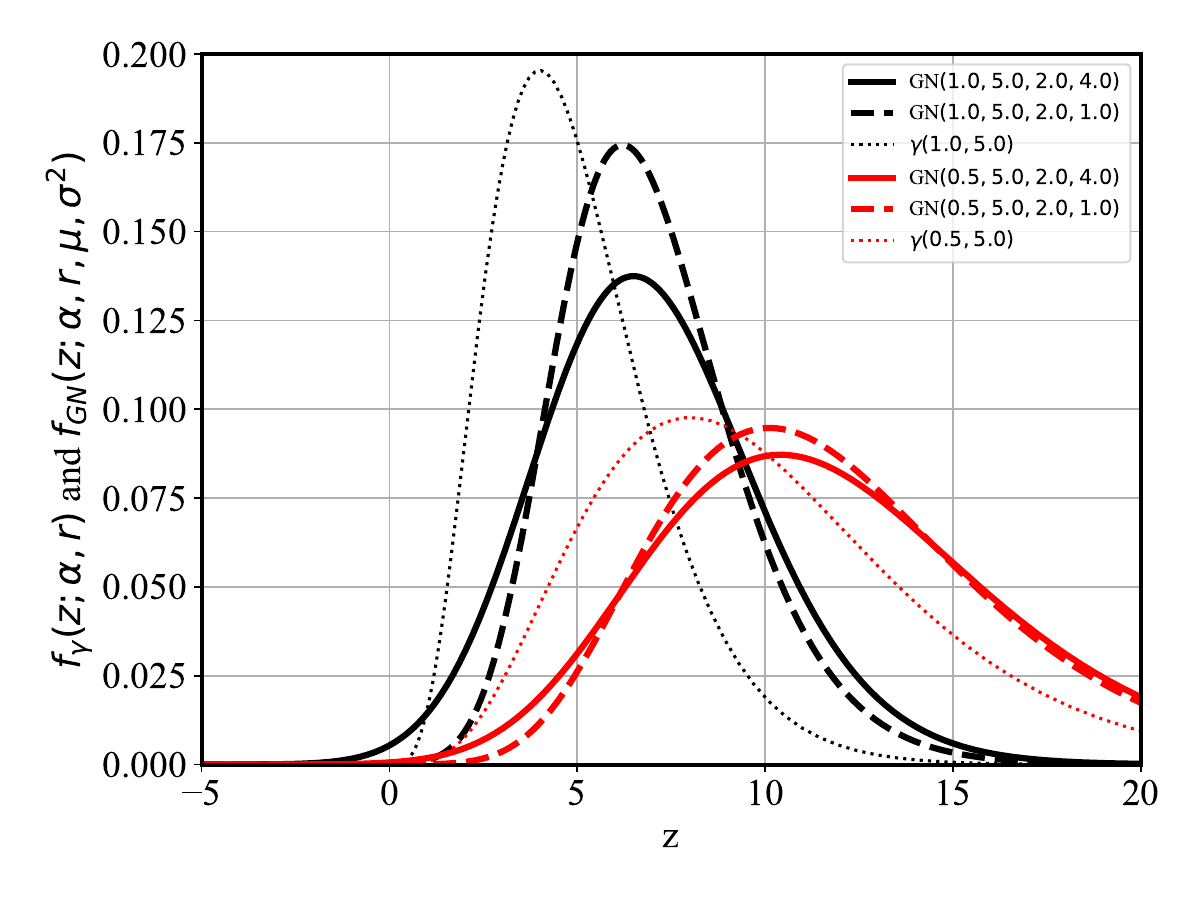}
    \caption{The \ng\ $\ngsymbol(\alpha,r,\mu,\sigma^2)$ distribution for selected values of the 
    the rate parameter of the $\gamma$ distribution ($\alpha=1, \nicefrac{1}{2}$) and of
    the variance of the normal distribution ($\sigma=2,1 $), and two reference values for
    the shape parameter of the gamma distribution ($r=5$) and the mean of the normal
    ($\mu=10$).}
    \label{fig:fGN}
\end{figure}

\begin{figure}
  \centering
  \includegraphics[width=4.2in]{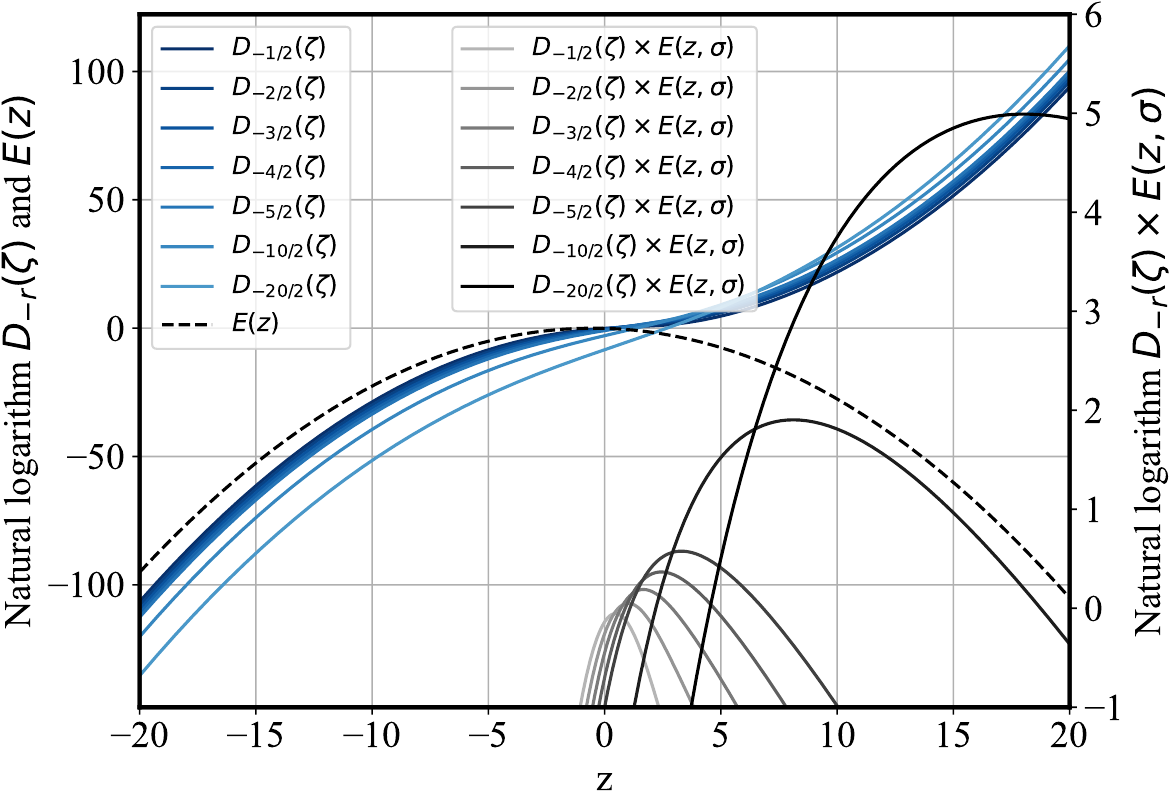}
  \caption{Parabolic cylinder functions 
  $D_{-r}(\zeta)$ as a function of the argument $z$, with $\zeta$ according to \eqref{eq:zeta},
  and for selected values of the shape parameter $r$.
  The mean of the normal was set to $\mu=0$,
  the variance to a fiducial value of $\sigma=1$, and the rate parameter 
  was also fixed at $\alpha=\nicefrac{1}{2}$. 
  On the right scale are values of the
  product of the parabolic cylinder function with the exponential function $E(z)$ 
  according to \eqref{eq:E},
  which is an unnormalized version of the pdf according to \eqref{eq:fNG}.}
  \label{fig:PCF}
\end{figure}

The pdf of a $\ngsymbol(\alpha, r, \mu,\sigma^2)$ random variable  can therefore be written in compact form as
\begin{equation}
  f_{\ngsymbol}(z; \alpha, r, \mu, \sigma^2) = \dfrac{(\alpha \sigma)^r}{\sqrt{2 \pi \sigma^2}} \;
    D_{-r}\left( \zeta \right) 
	\cdot E(z), 
	\label{eq:fNG}
\end{equation}
where
\begin{equation}
    \zeta =  \alpha \sigma + \dfrac{(\mu-z)}{\sigma}
    \label{eq:zeta}
\end{equation}
is the argument of the parabolic cylinder function, and $E(z)$ is an exponential function with
\begin{equation}
 \ln E(z)=
  -\dfrac{1}{4} \left( \sigma \alpha  + \tfrac{(z-\mu)}{\sigma}\right)^2 + \dfrac{\sigma^2 \alpha^2 }{2} 
   = \dfrac{\zeta^2}{4}  - \dfrac{(z-\mu)^2}{2 \sigma^2}.
  \label{eq:E}
\end{equation}
Equation~\ref{eq:fNG} is the most compact form for the pdf of the \ng\ distribution,
which can be evaluated via a parabolic cylinder function and elementary functions.
The general behavior of the pdf of the $\ngsymbol(\alpha,r,\mu,\sigma^2)$ is illustrated in
Fig.~\ref{fig:fGN}. Key features are a shift by $+\mu$ with respect to the $\gamma(\alpha,r)$ pdf (illustrated as a dotted curve), a broadening of the distribution, and a negative tail that is not present in the gamma distribution.
Figure~\ref{fig:PCF} further illustrates the behavior of the two functions $D_{-r}(\zeta)$
and $E(z)$ and of their product, which is proportional to the
pdf of the distribution according to \eqref{eq:fNG}. Given the large dynamical range of both functions, it is convenient
to work with the logarithms of the two functions.

For comparison, we report the form of the pdf of the \ng\ distribution as provided by \cite{plancade2012},
who first introduced it:
\begin{equation}
    f^{\textrm{ng}}_{\mu,\sigma,k,\theta}(x)=\int f_{k,\theta}^{\textrm{gam}} f^{\textrm{norm}}_{\mu,\sigma}(x-t) dt
\end{equation}
where $k$ is the shape parameter and $\theta$ the scale parameter of the gamma distribution. Their expression is equivalent to \eqref{eq:conv}, and the integral was evaluated numerically by way of a Fast Fourier Transform.

\subsection{Key properties}

\begin{property}[Translation property with respect to the $\mu$ parameter]
  \label{pr1}
The pdf of the \ng\ distribution has the property
\begin{equation}
    f_{\ngsymbol}(z-\mu; \alpha, r, \mu,\sigma^2) = f_{\ngsymbol}(z; \alpha, r, \mu=0, \sigma^2).
    \label{eq:shift}
\end{equation}
\end{property}
This property derives from the fact that the $\mu$ parameter 
  enters the pdf only as a function of $(z - \mu)$.
  This is the same translation property that the normal distribution has. The $\mu$ parameter
  is therefore simply a location parameter that does not otherwise affect the shape of the
  distribution.

\begin{property}[Mean, Variance, and Skewness] The mean, variance, and skewness of a 
$Z \sim \ngsymbol(\alpha,r,\mu,\sigma^2)$ variable are
\[
\begin{cases}
    \E[Z] = \mu+\dfrac{r}{\alpha}\\[10pt]
    \Var(Z) = \sigma^2 + \dfrac{r}{\alpha^2}\\
    \E\{[Z-E(Z)]^3\}=\dfrac{2}{\sqrt{r}}.
\end{cases}
\]
\label{pr2}
\end{property}
This property is an immediate consequence of the independence of the $X$ and $Y$ variables and the moments of the respective distributions.
\begin{property}[Large--mean normal approximation] 
For large values of $r/\alpha$, where $r$ is the shape parameter and $\alpha$ the rate parameter
of the gamma distribution, 
\[
  \ngsymbol \overset{a}{\sim} N((r/\alpha)+\mu, (r/\alpha^2)+\sigma^2),
\]
\label{pr3}
\end{property}
  This asymptotic convergence in distribution of 
  a $X \sim \gamma(\alpha, r)$ variable to a $N(r/\alpha, r/\alpha^2)$ 
  for large values of the mean $r/\alpha$ of the gamma distribution
  is a direct consequence of the
  central limit theorem, since a gamma distribution is the sum of $r$ independent 
  exponential random variables with finite mean and variance \citep[see, e.g.,][]{hogg2023}. 

\begin{property}[Closedness under convolution for fixed $\alpha$]
Let $Z_1$ and $Z_2$ be two independent \ng\ random variables with the same $\alpha$ parameter, i.e., $Z_1 \sim \ngsymbol(\alpha, r_1, \mu_1, \sigma_1^2)$ and $Z_2 \sim \ngsymbol(\alpha, r_2, \mu_2, \sigma_2^2)$. Then
\[
Z_1 + Z_2 \sim \ngsymbol(\alpha, r_1+r_2, \mu_1+\mu_2, \sigma_1^2+\sigma_2^2).
\]
    \label{pr4}
\end{property}
This property is an immediate consequence of the independence of the random variables, and the 
additive properties of the gamma and normal distributions. Specifically, if
$X_1 \sim \gamma(\alpha, r_1)$ and $X_2 \sim \gamma(\alpha, r_2)$ are independent,
then $X_1+X_2 \sim \gamma(\alpha, r_1+r_2)$; and if 
$Y_1 \sim N(\mu_1, \sigma_1^2)$ and $Y_2 \sim N(\mu_2, \sigma_2^2)$,  then $Y_1+Y_2 \sim N(\mu_1+\mu_2, \sigma^2_1+\sigma_2^2)$. This property extends to the sum of any number of
independent variables and it also means that, for a fixed $\alpha$, the \ng\ distribution is infinitely divisible.

\section{Special cases of the \ng\ distribution}
\label{sec:special}
This section describes special cases of the \ng\ distribution and their properties, for certain choices of the four parameters, that are of common use in probability and
in statistical applications.

\subsection{The \en\ distribution}
\label{sec:en}
When the shape parameter of the gamma distribution is $r=1$, the gamma distribution becomes an exponential distribution with rate parameter 
$\alpha$ (or scale parameter $1/\alpha$). For $r=1$, the pdf of the \ng\ distribution
can be simplified making use of
\begin{equation}
  D_{-1}(z) = e^{z^2/4} \sqrt{\dfrac{\pi}{2}} \left(1 - \erf\left(\dfrac{z}{\sqrt{2}}\right) \right)
\end{equation}
where $\erf(x)$ is the error function \citep[see 9.254.1 of][ and App.~\ref{sec:appA}]{gradshteyn2007}. 

A random variable $Z \sim \ngsymbol(r=1,\alpha,\mu,\sigma^2)$ is therefore
said to be an \en\ or exponential--Gaussian variable $Z \sim \ensymbol(\alpha,\mu,\sigma^2)$, with probability 
distribution function
\begin{equation}
    f_{\ensymbol}(z) = \dfrac{\alpha}{2} \left(1-\erf(\zeta/\sqrt{2})\right) e^{\zeta^2/4} E(z),
    \label{eq:fEG}
\end{equation}
where
\begin{equation}
    \begin{cases}
        \zeta = \alpha \sigma + \dfrac{\mu - z }{\sigma}\\[10pt]
        \ln E(z) + \dfrac{\zeta^2}{4} = \dfrac{\alpha}{2}\left( 2(\mu -z) + \sigma^2 \alpha \right).
    \end{cases}
\end{equation}
This distribution is the same exponential--normal distribution
that has been used for chromatography and other applications \cite[see, e.g.][]{gruskha1972, delley1985}.
The pdf in Equation~\ref{eq:fEG} is equivalent to the one used by various authors in this
field, including \cite{kalambet2011, xie2009}.

\begin{figure}
    \centering
    \includegraphics[width=4.2in]{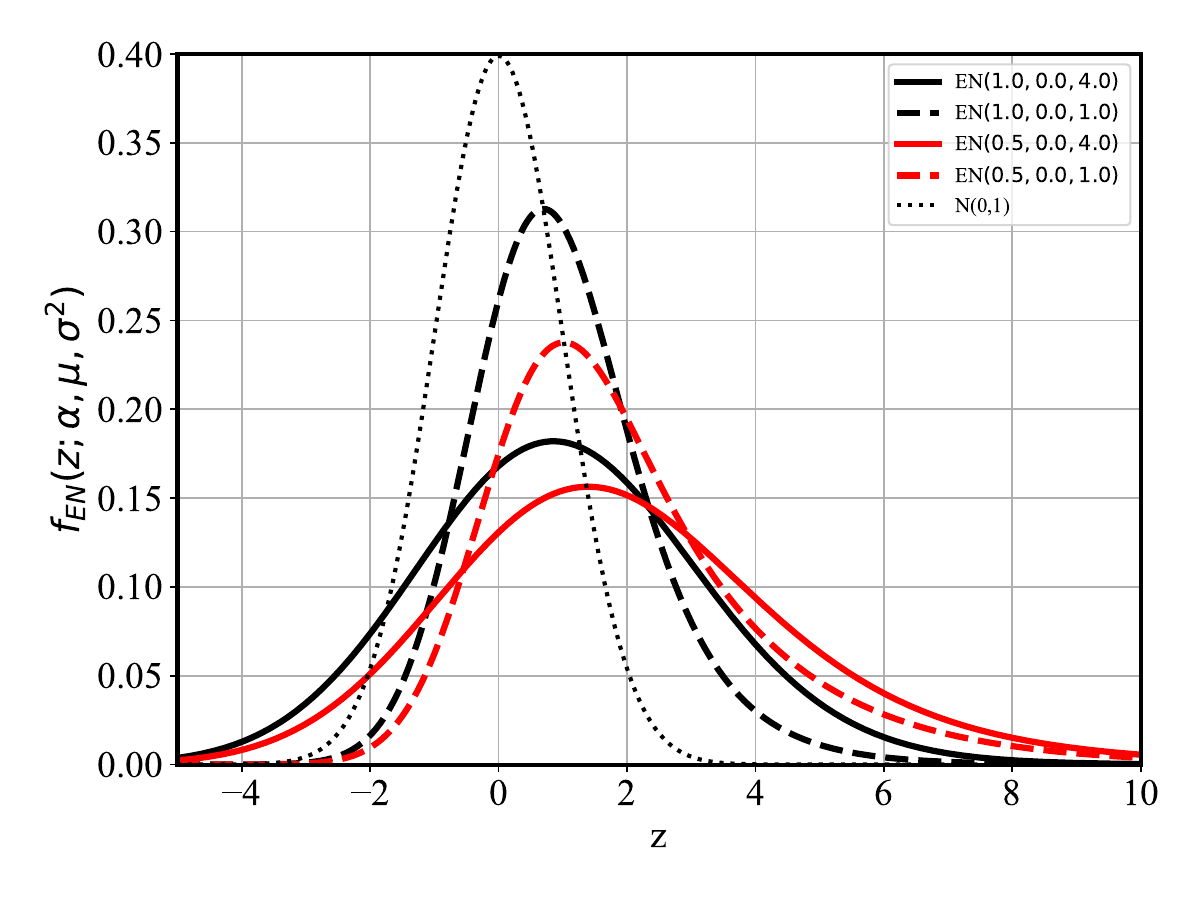}
    \caption{Probability density function of an \en\ variable $\ensymbol(\alpha,\mu,\sigma^2)$, for a value $\mu=0$ and selected values of the $\alpha$ and $\sigma^2$
    parameters.}
    \label{fig:fEN}
\end{figure}

Figure~\ref{fig:fEN} illustrates the behavior of the pdf of the \en\ distribution, which
features  a mean $\E[Z]=\mu + 1/\alpha$ and a variance $\Var(Z) = \sigma^2+ 1/\alpha^2$.
The effect of the convolution  with an exponential is to increase both the mean and variance of the
normal distribution. Additional properties were studied in \cite{gruskha1972}.

\subsection{The chi--squared--normal or overdispersed chi--squared distribution}
\label{sec:odchi}
When $\alpha=\nicefrac{1}{2}$ and we set $r=\nicefrac{\nu}{2}$,
the gamma distribution $\gamma(\alpha, r)$ becomes a chi--squared distribution 
$\chi^2(\nu)$ with degrees of freedom parameter $\nu$. 
The chi--squared distribution \citep[e.g.][]{helmert1876, fisher1924} has occupied a central role 
in statistics since K.~Pearson's 
introduction of the $\chi^2$ test of goodness of fit \citep{pearson1900}. One of its uses in statistics is
as the parent distribution of a number
of goodness--of--fit statistics that are in common use.
One of the most common uses is for normal data,
whereas chi--squared is the parent distribution of the Gaussian log--likelihood statistic 
obtained from maximum--likelihood optimization
\citep[often referred to as the  $\chi^2$ or $S$ statistic, e.g.][]{fisher1925}. In many statistical
applications,  $\nu \in\mathbb{N}$ is an integer number of degrees of freedom of the distribution, but the
properties derived in this paper apply to the more general case of a positive real $\nu$ parameter.

A random variable $Z \sim \ngsymbol(r=\nicefrac{\nu}{2},\alpha=\nicefrac{1}{2}, \mu, \sigma^2)$ is said to be a chi--squared--normal or an \odchi\ variable $Z \sim \odchisymbol(\nu, \mu, \sigma^2)$.
Its probability distribution function is given, according to \eqref{eq:fNG}, by
\begin{equation}
  f_{B}(z; \nu, \mu, \sigma^2) = \dfrac{1}{2^{\nu/2} \sqrt{2 \pi}\; \sigma^{1-\nu/2}} D_{-\nu/2}\left( \zeta \right) 
	\cdot E(z,\sigma), 
	\label{eq:fB}
\end{equation}
where
\begin{equation}
\begin{cases}
    \zeta = \dfrac{\sigma}{2} + \dfrac{(\mu-z)}{\sigma} \\[10pt]
     \ln E(z,\sigma)= -\left( \tfrac{\sigma^2 + 2(z-\mu)}{4 \sigma}\right)^2 + \dfrac{\sigma^2}{8}.
    \end{cases}
\end{equation}
The mean is $\E[Z]=\mu+\nu$, and the variance is $\Var(Z)=2 \nu + \sigma^2$. 

\begin{figure}[!t]
  \centering
  \includegraphics[width=4.2in]{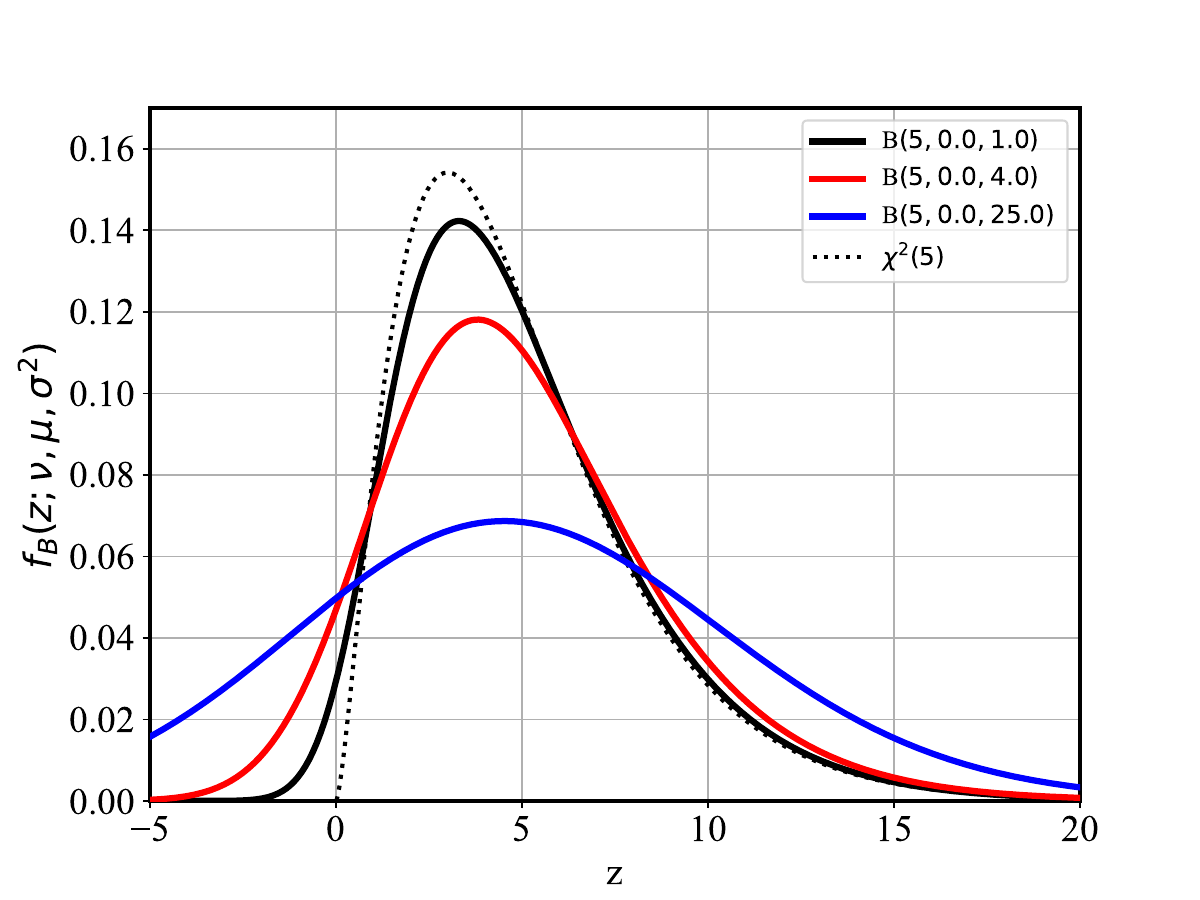}
  \caption{Probability density function of the \odchi\ variable $B(\nu,\mu=0,\sigma^2)$ for a fiducial value of the
  number of degrees of freedom, and as a function of the overdispersion parameter $\sigma^2$.}
  \label{fig:fBDSigma}
\end{figure}

\begin{figure}
  \centering
  \includegraphics[width=4.2in]{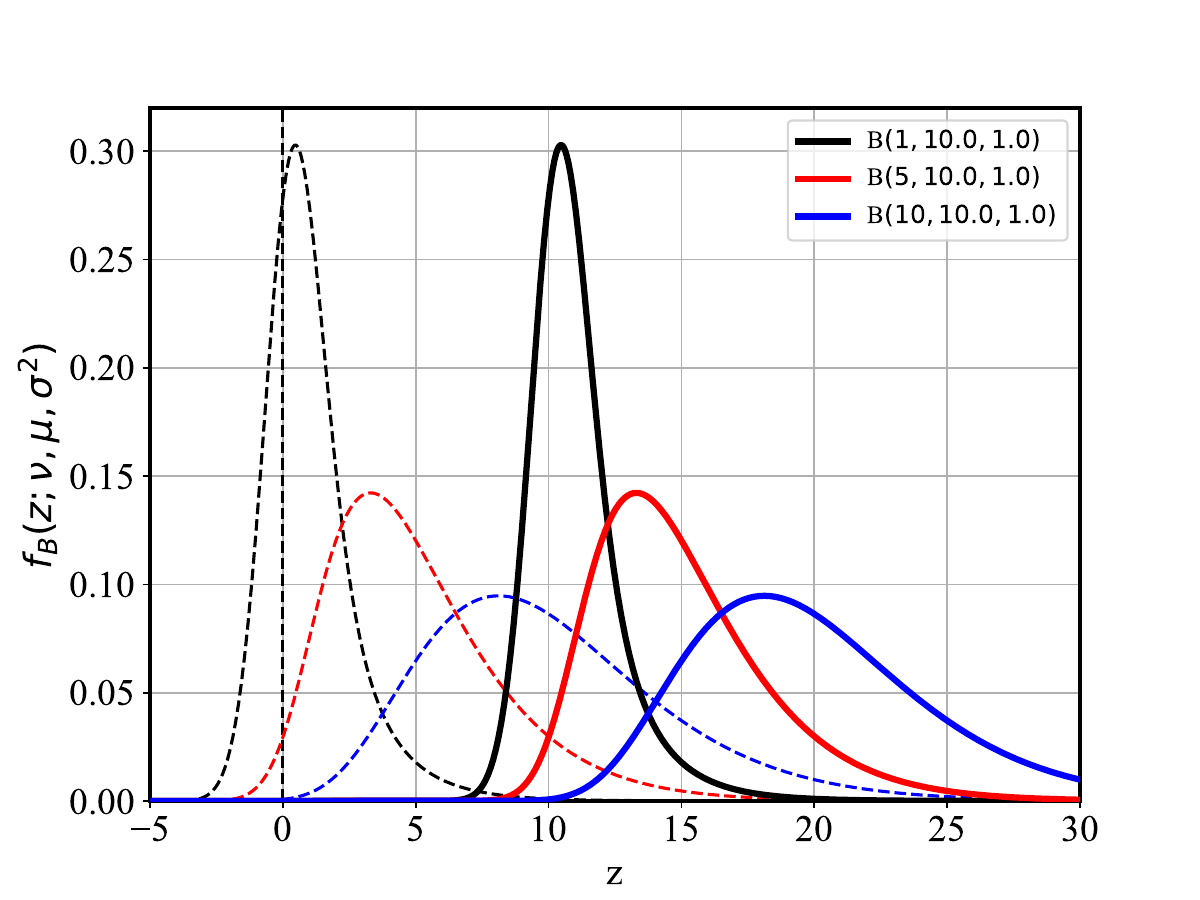}
\caption{Probability density function of the \odchi\ variable $B(\nu,\mu,\sigma^2)$ for fiducial
  values of the $\mu$ and $\sigma^2$ parameters, and for selected values of the
  number of degrees of freedom $\nu$. The bias parameter $\mu$ only provides a shift of the distributions by a fixed amount $+\mu$ relative to the
corresponding $\mu=0$ distributions, which are plotted as dashed curves.}
  \label{fig:fBDmu}
\end{figure}

Figure~\ref{fig:fBDSigma} illustrates the pdf of a $B(\nu,\mu,\sigma^2)$ variable for $\mu=0$ and a fixed value of $\nu$,
as a function of the overdispersion parameter $\sigma^2$. The broadening of the distribution 
with increasing $\sigma^2$ is the reason for the name \emph{overdispersed} $\chi^2$ distribution,
and it leads to a tail of negative values that is not present in the $\chi^2(\nu)$ distribution. 
Figure~\ref{fig:fBDmu} illustrates the behavior of the $B(\nu,\mu,\sigma^2)$ distribution as a function
of the number of degrees of freedom, for a fiducial value of the $\mu$ and $\sigma^2$ parameters.  The corresponding $B(\nu,\mu=0,\sigma^2)$ are also plotted as dashed curves to illustrate the translation property (Property~\ref{pr1}). The overdispersed chi--squared
distribution was first introduced by 
\cite{bonamente2023} 
for the case of $\mu=0$, and its density is now provided in full form in \eqref{eq:fB}.

\begin{property}[Closedness over convolution of the \odchi\ distribution] Let $Z_1 \sim \odchisymbol(\nu_1, \mu_1, \sigma^2_1)$ and $Z_2 \sim \odchisymbol(\nu_2, \mu_2, \sigma^2_2)$
be two independent \odchi\ distribution. Then
\[
Z_1+Z_2 \sim \odchisymbol(\nu_1+\nu_2, \mu_1+\mu_2, \sigma^2_1+\sigma^2_2)
\]
\label{pr5}    
\end{property}
This is an immediate consequence of Property~\eqref{pr4}, and it applies to the sum of any number
of independent \odchi\ variables. This means that the \odchi\ distribution is infinitely divisible.

\begin{table}
  \footnotesize
  \caption{Selected percentiles or one--sided upper critical values of the $B(\nu,\mu=0,\sigma^2 )$ family of distributions for $\nu=1,2,3,4,5$ and 10,
  and for representative values of the overdispersion parameter $\sigma^2$. The $\chi^2$ entries correspond to
  the case of no overdispersion.}
  \label{tab:ODC}
  \begin{tabular}{l|rrrrrr} 
    \hline
$\sigma$ & \multicolumn{6}{c}{100$p$th percentile} \\
\hline
	& \multicolumn{6}{c}{$p=0.5$ (median)} \\ 
	& \multicolumn{6}{c}{Number of d.o.f. ($\nu$)}  \\
      & 1 & 2 & 3 & 4 & 5 & 10 \\ 
      & \multicolumn{6}{c}{\hrulefill} \\
    1 & 0.738& 1.577 & 2.492 & 3.448 & 4.423& 9.376 \\ 
    2 & 0.859& 1.752 & 2.671 & 3.612 & 4.568& 9.462 \\ 
    5 & 0.959& 1.922 & 2.888 & 3.857 & 4.829& 9.722 \\ 
    10& 0.988& 1.976 & 2.964 & 3.953 & 4.942& 9.894 \\ 
    $\chi^2$ &0.455& 1.386 & 2.366 & 3.357 & 4.351&  9.342\\ 
    \hline
        & \multicolumn{6}{c}{$p=0.90$} \\
	& \multicolumn{6}{c}{Number of d.o.f. ($\nu$)}  \\
	& 1 & 2 & 3 & 4 & 5 & 10 \\
	& \multicolumn{6}{c}{\hrulefill} \\
	1 & 3.067 & 4.855 & 6.460 & 7.964 & 9.404 & 16.111 \\
	2 & 4.061 & 5.584 & 7.066& 8.499 & 9.891 & 16.475\\
	5 & 7.659 & 8.907 & 10.149& 11.386 &12.617 & 18.688\\
	10&13.947 & 5.078 & 16.207& 17.335 & 18.462 &24.081\\
	$\chi^2$ & 2.706 & 4.605 & 6.251 & 7.779 & 9.236 & 15.987\\
	\hline
	& \multicolumn{6}{c}{$p=0.95$} \\ 
     & \multicolumn{6}{c}{Number of d.o.f. ($\nu$)}  \\
      & 1 & 2 & 3 & 4 & 5 & 10 \\ 
      & \multicolumn{6}{c}{\hrulefill} \\
    1 &4.163& 6.241 & 8.032 & 9.684 & 11.252 & 18.447 \\ 
    2 &5.171& 6.988 & 8.673 & 10.262 &11.785 & 18.861 \\ 
    5 &9.603& 10.969& 12.322 & 13.662&14.990 & 21.469 \\ 
    10&17.632& 18.813&19.991 & 21.168&22.343 & 28.190 \\ 
    $\chi^2$ & 3.841& 5.991 & 7.815 &9.488 &11.070 &18.307 \\ 
    \hline
    & \multicolumn{6}{c}{$p=0.99$} \\
    & \multicolumn{6}{c}{Number of d.o.f. ($\nu$)}  \\
      & 1 & 2 & 3 & 4 & 5 & 10 \\
      & \multicolumn{6}{c}{\hrulefill} \\
    1 & 6.924 & 9.460 & 11.573 & 13.489 & 15.286 & 23.373 \\
    2 & 7.817 &10.210& 12.251 &14.120 & 15.880 & 23.859\\
    5 & 13.349 & 15.012 & 16.627 & 18.200 &19.738 & 27.055\\
    10 & 24.562 & 25.855 & 27.143 & 28.426 &29.705 & 36.036\\
    $\chi^2$ & 6.635 & 9.210 & 11.345 & 13.277 & 15.086 & 23.209\\
    \hline
    & \multicolumn{6}{c}{$p=0.999$} \\
    & \multicolumn{6}{c}{Number of d.o.f. ($\nu$)}  \\
    & 1 & 2 & 3 & 4 & 5 & 10 \\
    & \multicolumn{6}{c}{\hrulefill} \\
    1 &11.101& 14.066 & 16.501 & 18.691 & 20.729 & 29.771 \\
    2 & 11.927& 14.816 & 17.203 & 19.356& 21.366 & 30.316 \\
    5 & 17.839&  19.992& 21.984 & 23.866& 25.669 & 33.984 \\
    10 &  32.371 & 33.826&35.269 & 36.701 &38.123 &45.099  \\
    $\chi^2$ & 10.827 &13.816  &16.266 & 18.467& 20.515& 29.588  \\
\hline
  \end{tabular}
\end{table}

\section{Maximum likelihood estimators}
\label{sec:ML}
The compact form for the pdf of the \ng\ distribution enables analytic expressions for
the maximum--likelihood score equations and  the Fisher information matrix. Detailed
calculations are provided in in App.~\ref{sec:appB}, and the main results are summarized and discussed in this section. 

\subsection{Score equations and information matrix of the \ng\ distribution}
With $\theta=(\alpha,r,\mu,\sigma)$ and for $N$ iid measurements $z_i \sim \ngsymbol(\theta)$, the score equations are evaluated from the log--likelihood derivatives derived in \eqref{eq:score}. 
Summing over the $N$ measurements leads to the following first--order derivatives
of the log--likelihood $\mathcal{L}$, and the associated score equations for the four--parameter \ng\ distribution:\begin{equation}
    \begin{cases}
     \dfrac{\partial \mathcal{L}}{\partial \alpha} = N\left(\dfrac{r}{\alpha} + S \cdot \sigma + \dfrac{\alpha \sigma^2}{2} + \dfrac{Z_1}{2} \right)=0\, \dotfill \text{(Score eq. for $\alpha$)}\\[10pt]
      \dfrac{\partial \mathcal{L}}{\partial r} = N\left( \ln (\alpha \sigma) + T -  \psi(r) \right)= 0\, \dotfill \text{(Score eq. for $r$)} \\[10pt]
        
       \dfrac{\partial \mathcal{L}}{\partial \mu} =N\left( \dfrac{S}{\sigma} + \dfrac{\alpha}{2} - \dfrac{Z_1}{2 \sigma^2} \right)=0\,  \dotfill \text{(Score eq. for $\mu$)} \\[10pt]
    \dfrac{\partial \mathcal{L}}{\partial \sigma} =N\left(\dfrac{r-1}{\sigma} +S \cdot \alpha -\dfrac{S_{\mu}}{\sigma^2} + \dfrac{\sigma \alpha^2}{2}+\dfrac{Z_2}{2 \sigma^3} \right)=0\,  \dots \text{(Score eq. for $\sigma$)}
    \end{cases}
    \label{eq:scoreShort}
\end{equation}
where the relevant sums $S$, $T$, $Z_1$, $Z_2$ and $S_{\mu}$ are defined in App.~\ref{sec:appB}. 
This is
a system of four nonlinear equations that must be solved for $\hat{\theta}$, the maximum likelihood estimate of $\theta$. The second--order derivatives
of the functions in \eqref{eq:scoreShort} are used for the information matrix.

A numerical solution of these equations
requires the evaluation of the parabolic cylinder function $D_{-r}(\zeta_i)$, with $\zeta_i=\zeta(z_i)$
according to \eqref{eq:zeta}, its derivative with respect to $\zeta$, which is a function
of parabolic cylinder functions according to the recursive equation \eqref{eq:Dpprime},
its derivative with respect of the index $r$ which is given by \eqref{eq:Ar},
the digamma
function $\psi(r)$ which is the logarithmic derivative of the $\Gamma(r)$ function, plus elementary functions.

It has been pointed out by \cite{sprott1983} that the maximum likelihood estimates of certain convolution
distributions, which include variables in the exponential family such as the ones considered in this paper, feature the additional likelihood equation  $\E[Z] = \overline{z}$. 
Although \cite{sprott1983} only provides a proof of this property
for the case of the convolution of two
one--parameter distributions such as the Poisson and the binomial, we can immediately see 
that the combination of the score equations for $\alpha$ and $\mu$ leads to
\begin{equation}
    \mu + \dfrac{r}{\alpha} = \overline{z} = \E[Z].
\end{equation} 
Hereafter this relationship will be referred to as  \textit{Sprott's equation} for the \ng\ distribution.

The use of such an additional equation was envisioned by \cite{sprott1983} as a means to
simplify the process of \ml\ estimation, by replacing one of the score equations with the
typically simpler Sprott's equation. In our case, the use of Sprott's equation
shows that the two score equations for $\alpha$ and $\mu$
become identical for all values of $\theta$ and of $z$. Accordingly, it is easy to show that \eqref{eq:scoreShort}
is reduced to a system of three equations, plus
Sprott's equation,
\begin{equation}
    \begin{cases}
        \ln (\alpha \sigma) + T  -  \psi(r) = 0\, \dotfill \text{(Score eq. for $r$)} \\[10pt]
        \dfrac{r}{2 \alpha} + S \cdot \sigma + \dfrac{\alpha \sigma^2}{2} = 0\, \dots 
        \dotfill \text{(Score eq. for $\alpha$ or $\mu$)}\\[10pt]
        \dfrac{r}{2} -1  +  \dfrac{Z_2}{2 \sigma^2} -\dfrac{S_{\mu}}{\sigma} = 0\,  \dotfill \text{(Score eq. for $\sigma$)} \\[10pt] 
        r = (\overline{z} - \mu) \alpha\, \dotfill \text{(Sprott's equation)}.
    \end{cases}
    \label{eq:scoreShortSprott}
\end{equation}

For the observed Fisher information matrix $I$, defined by
\begin{equation}
\begin{aligned}
    I=- &\left( \sum\limits_{i=1}^N \dfrac{\partial}{\partial \bm{\theta}} 
    \left[\dfrac{\partial}{\partial \alpha} \ln f_{\ngsymbol}(z_i)\right],
    \sum\limits_{i=1}^N \dfrac{\partial}{\partial \bm{\theta}} 
    \left[\dfrac{\partial}{\partial r} \ln f_{\ngsymbol}(z_i)\right]\right.,\\
    & \left. 
     \sum\limits_{i=1}^N \dfrac{\partial}{\partial \bm{\theta}} 
    \left[\dfrac{\partial}{\partial \mu} \ln f_{\ngsymbol}(z_i)\right],
    \sum\limits_{i=1}^N \dfrac{\partial}{\partial \bm{\theta}} 
    \left[\dfrac{\partial}{\partial \sigma} \ln f_{\ngsymbol}(z_i)\right]
    \right)=  (I_1^T, I_2^T, I_3^T, I_4^T),
\end{aligned}
\end{equation}
the second--order derivatives of the log--likelihood 
are carried out in App.~\ref{sec:appB}, see \eqref{eq:I1} through \eqref{eq:I4}. These equations
 can be used to evaluate the the observed information matrix by summing over the $N$ measurements, leading to the following  symmetric matrix:
\begin{equation}
\begin{cases}
    I_1/N=-\left( -\dfrac{r}{\alpha^2} + \sigma^2 S_{\zeta} + \dfrac{\sigma^2}{2},
    \dfrac{1}{\alpha} + \sigma S_r,  S_{\zeta} + \dfrac{1}{2}, 
    \alpha \sigma S_{\zeta}  - \dfrac{S_{\zeta, \mu}}{\sigma}  + S+\alpha \sigma\right)\\[10pt]
    I_2/N=-\left(\dots, T_r - \psi_1(r),  \dfrac{S_r}{\sigma}, \dfrac{1}{\sigma} + \alpha S_r 
    -\dfrac{S_{r, \mu}}{\sigma^2}  \right)\\[10pt]
    I_3/N=-\left( \dots, \dots,  \dfrac{S_{\zeta}}{\sigma^2}- \dfrac{1}{2 \sigma^2}, 
     \dfrac{\alpha S_{\zeta}}{\sigma} - \dfrac{S}{\sigma^2} - \dfrac{S_{\zeta, \mu}}{\sigma^3} 
     + \dfrac{Z_1}{\sigma^3} \right)\\[10pt]
    I_4/N=-\left(\dots, \dots,\dots, -\dfrac{r-1}{\sigma^2}+\alpha^2 S_{\zeta} 
    + \dfrac{S_{\zeta,\mu^2}}{\sigma^4} 
    - \dfrac{ 2 \alpha S_{\zeta,\mu}}{\sigma^2}  + \dfrac{2 S_{\mu}}{\sigma^3}  +\dfrac{\alpha^2}{2}
    -\dfrac{3 Z_2}{2 \sigma^4} \right).
\end{cases}
\label{eq:I1text}
\end{equation}
All relevant sums (e.g, $S_{\zeta}$, $S_r$ etc.) are defined in App.~\ref{sec:appB}.
Finally, the asymptotic covariance matrix  $\hat{\epsilon} = \hat{I}^{-1}$ simply requires the inversion 
of the $4 \times 4$ matrix \eqref{eq:I1text}.

\subsection{Parameter identifiability}
\label{sec:identifiability}
Equations~\eqref{eq:scoreShort}, or equivalently \eqref{eq:scoreShortSprott},
yield the maximum likelihood estimates $\hat{\theta}$ of the four parameters of the \ng\ distribution, provided that all parameters are indeed \textit{identifiable}. 

The problem of identification in parametric models is related to the existence of a solution of the estimating equations for a given parameter
\citep[e.g., Sec.~29.11, ][]{kendall1979}. This problem has received
much attention in the statistics literature, especially for econometric
models \citep[e.g.,][]{wald1950, fisher1966, amemiya1985}. In particular, T. Rothenberg  has shown that a necessary and sufficient condition for the
{identifiability} of a parameter set $\theta_0$ is that the Fisher information matrix $R_0=\E[I_0]$
is non--singular \citep[see Theorem~1 in][]{rothenberg1971}. This result, hereafter referred
to as Rothenberg's theorem, will be used to address the identifiability of the parameters
in the \ng\ model.

Evaluation of the expectation of the observed information matrix \eqref{eq:I1text} is complicated by the
non--trivial integrals in the sums \eqref{eq:SumsI} contained in the matrix.
In place of the expectation $\E[I]$, we use the observed information matrix by drawing a large number $N$ of
samples from the parent distribution, and assess the identification of parameters for a
number of representative cases using this large--$N$ observed information matrix. 
In the asymptotic limit 
of large $N$, the observed information matrix converges to the
expected information matrix by the independence of the measurements
and the law of large numbers, and therefore the Rothenberg theorem
can be used in an approximate form for a finite sample, thus overcoming the difficulties associated with the exact evaluation of the expectations of \eqref{eq:I1text}.
The results of these numerical tests are provided next.

\subsection{Numerical tests}

In all cases, we draw $N=100$ random samples from the parent distribution for a fixed
set of representative values. In particular, we chose two representative cases as $\theta=(\nicefrac{1}{2},\nicefrac{1}{2},5,1)$ with $\alpha=\nicefrac{1}{2}$ (which also applies to the \odchi\ distribution)
and $(\nicefrac{1}{2},1,1,1)$ with $r=1$ (which is the simplified case of the
\en\ distribution). We also test other parameter sets, as needed, to further explore parameter space. 
We then
solve the score equations \eqref{eq:scoreShort}, and evaluate the observed
information matrix \eqref{eq:I1text}. Additional details on the numerical tests
are provided in App.~\ref{sec:appC}, and the main results are summarized in the following.

\subsubsection{The full four--parameter \ng\ distribution}
\label{sec:4paramFit}
We find evidence that the information matrix is singular or nearly--singular,
 when all parameters are estimated simultaneously from \eqref{eq:scoreShort}. In fact, for all simulations that were run, (a) one of the four eigenvalues of $I$ is much smaller than the others and  numerically close to zero (relative to the number $N$); (b) the determinant of the observed information matrix has a small value, relative to $N^4$, and (c) the determinant is sometimes negative, indicating that
the observed information matrix is not positive--definite, as it ought to be.

This empirical evidence suggests that $\rank(I)<4$, although an exact proof
would require the evaluation of the expectation of $I$, which we do not attempt in this paper.
If true, this would imply that only three parameter combinations can be estimated from
a given dataset. The pdf of the \ng\ distribution itself 
(see Sec.~\ref{sec:ng}) does not appear to depend on just three parameter combinations, and further considerations are needed to identify possible parameter combinations that result in a $3 \times 3$ non--singular information matrix. 

This tentative finding is not in contradiction with the successful \ml\ estimation of the
\ng\ distribution performed by \cite{plancade2012}, who first introduced the \ng\ distribution. In fact, their method of estimation of the 
four parameters relies on two independent sets of data, one of which is for a
background component that is $N(\mu, \sigma^2)$ distributed. The use of an additional set of data to estimate two of the four parameters clearly provided the additional information
that results in the simultaneous determination of all parameters.  But in many applications, such additional data will not be available.

Accordingly, we proceed with testing the score equations and the
information matrix derived in this paper for the two three--parameter 
special cases of the \ng\ distribution 
discussed in Sec~\ref{sec:special}: the \odchi\ or chi--squared--normal distribution
$B(\nu,\mu,\sigma^2)$ that is obtained as a special case of the \ng\ distribution for a fixed value
of $\alpha=\nicefrac{1}{2}$ and with $\nu=2\,r$ (see Sec.~\ref{sec:odchi}); and for the
\en\ distribution that is obtained by fixing $r=1$ (see Sec~\ref{sec:en}).

\subsubsection{The \odchi\ or chi--squared--normal distribution}

Maximum likelihood estimation of the \odchi\ distribution $B(\nu,\mu,\sigma^2)$ is obtained 
from the usual sets of equations \eqref{eq:scoreShort} and \eqref{eq:I1text}, by
removing respectively the score equation for the fixed $\alpha=\nicefrac{1}{2}$ parameter
and the corresponding row and column in the information matrix. For all simulated data,
we consistently observed that the information matrix is positive definite, and that
the maximum likelihood estimates are consistent with the input parameters. Therefore, we find
empirical evidence that we can estimate the three parameters $\nu=2 r$, $\mu$ and
$\sigma^2$ well, even from a sample of moderate size ($N=100$). 

\begin{figure*}
    \centering
    \includegraphics[width=3.6in]{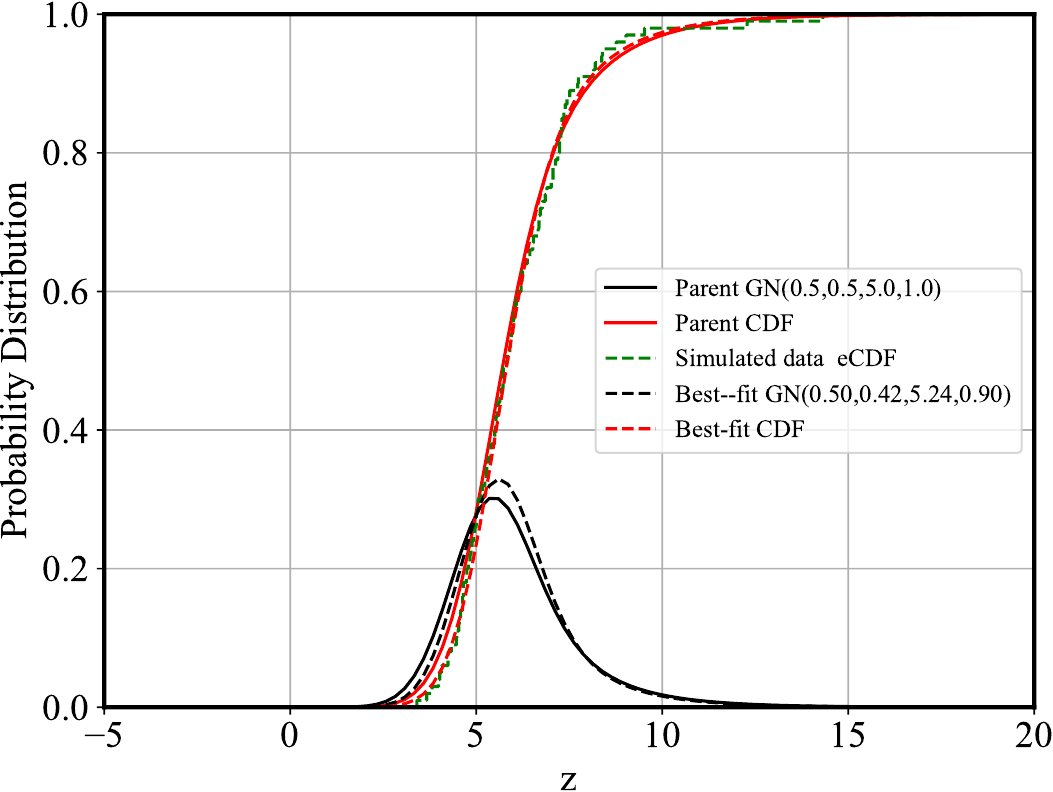}
     \includegraphics[width=3.6in]{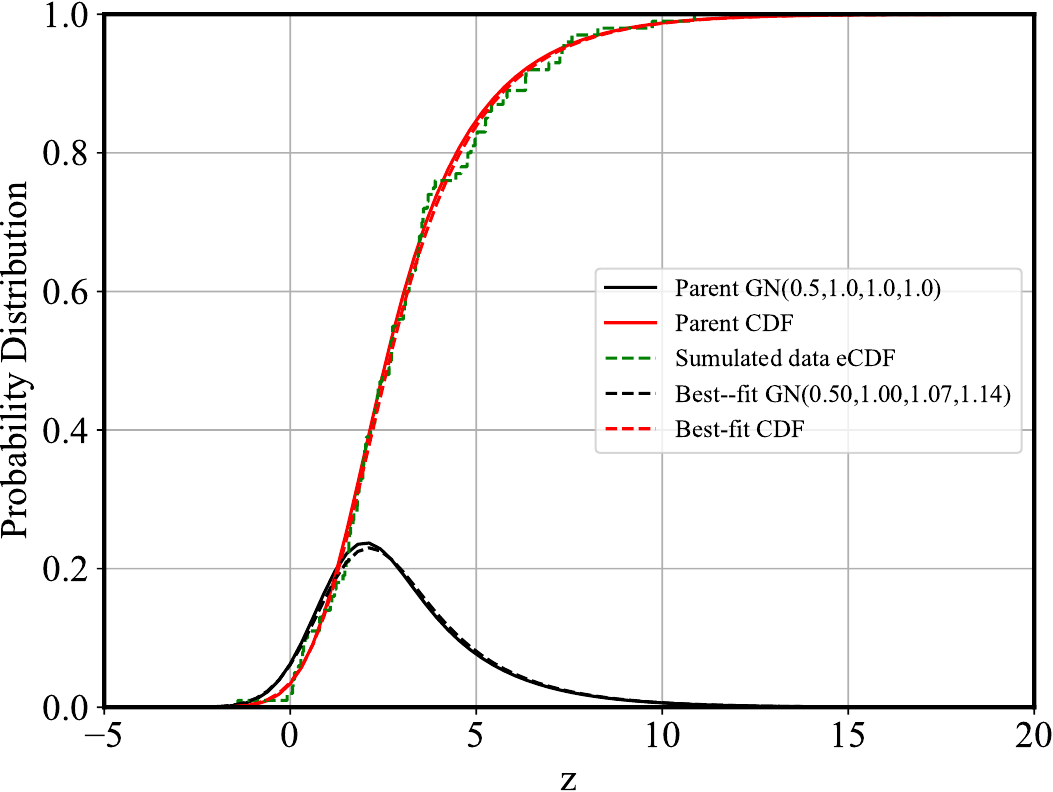}
    \caption{Top: \ml\ estimation of parameters for a \odchi\ distribution with
    fixed $\alpha=\nicefrac{1}{2}$; Bottom: for 
    an \en\ distribution with fixed $r=1$.}
    \label{fig:testE}
    \label{fig:testD}
\end{figure*}

An example of the results of the fit to a simulated dataset is provided
in the top panel of Fig.~\ref{fig:testE}.
The score equations yield
parameter estimates $(\hat{r},\hat{\mu},\hat{\sigma})=(0.42,5.24,0.95)$, with observed information matrix
\[
I = \begin{bmatrix}
    168.65 & 83.67 & 21.47 \\
    83.67 & 63.51 & -25.44\\
    21.47 & -25.44 & 88.77\\
\end{bmatrix}
\]
that has eigenvalues $(215.24,4.57, 101.12)$ and $\det(I)=99415.99$. Inversion of this
matrix leads to the
the observed covariance matrix
\[ 
\epsilon = \begin{bmatrix}
    0.0502 &  -0.0802 &   -0.0351\\
 -0.0802 &    0.1459 &   0.0612 \\
 -0.0351 &   0.0612 &   0.0373 \\
\end{bmatrix}
\]
Similar values are obtained for other equivalent simulations. 

It is useful to point out the  negative correlation that is typically observed between estimates of the $r$ and $\mu$ parameters, 
and between estimates of the $r$ and $\sigma$ parameters. These correlations are a direct result of, respectively,
the expectation and variance of the $Z$ variable, see Property~\ref{pr2}.
For example, as $r$ increases,
$\mu$ must decrease (for a fixed $\alpha$) to  maintain the same parent mean; and likewise as
$r$ increases, $\sigma^2$ must be reduced in value to keep the same parent variance. These negative correlations are therefore expected to apply to all applications in which the three parameters
$r, \mu, \sigma^2$ are simultaneously estimated from a single set of data.

In applications where the $\alpha=\nicefrac{1}{2}$ gamma distribution becomes a chi--squared
distribution with $\nu=2 r \in \mathbb{N}$, it is no longer meaningful to estimate the $r$ parameter
via the equations presented in this paper, which assumes the more general case of $r \in \mathbb{R}^+$. Instead, the $\nu$ parameter must be held fixed, and
a two--parameter estimation can be performed using only the two equations that apply to the
$\mu$ and $\sigma$ parameters.





\subsubsection{The \en\ distribution}

When $r=1$, the \ng\ distribution becomes the exponential--normal distribution 
discussed in Sec.~\ref{sec:en}. An example of parameter estimation 
is shown in the bottom panel of Fig.~\ref{fig:testD}, featuring an observed information matrix

\[
I =  
\begin{bmatrix}
    326.92 &  -61.54 &  -16.01\\
    -61.54 &   33.68 &  -14.02\\
   -16.01  & -14.02  &  42.63\\
\end{bmatrix}
\]
with eigenvalues $(339.87,11.86,51.50)$ and $\det(I)=207530.90$. The 
observed covariance
matrix is
\[
\epsilon = \begin{bmatrix}
  0.0060 & 0.0137 &  0.0068\\
  0.0137 & 0.0660 & 0.0268\\
  0.0068 &  0.0268 &  0.0348\\
\end{bmatrix}
\]
Since $r$ is fixed, there is now a positive correlation 
between the $(\alpha$, $\mu)$ and the $(\mu$, $\sigma^2)$ pairs of parameters. This is again
in accordance with the mean and variance of the $Z$ variable, as already discussed for the \odchi\ distribution above.

\section{Discussion of certain applications of the \ng\  distribution}
\label{sec:application}
The \ng\ distribution may apply to any statistic or random variable 
that is the sum of independent gamma and normal random variables (one of each), according to its
definition \eqref{eq:fNG}. It is therefore a general--purpose distribution that is useful for
a variety of applications, and it goes beyond the scope of this paper
to review all of them. In the following we briefly review and discuss 
a few possible {classes of} applications. {In particular, we
emphasize two categories
that are of interest to recent applications, and we perform 
an analysis on real data using the methods presented in this paper.}

\subsection{Noise--signal deconvolution}
\label{sec:applications1}

\subsubsection{General considerations}
The \ng\ distribution was proposed by \cite{plancade2012} as an extension
of the \en\ distribution \citep[e.g.][]{gruskha1972,xie2009} for the purpose of noise--signal decomposition in the analysis of fluorescence data from biological samples with the
\texttt{Illumina BeadArray} technology.
In that application, $Z=X+Y$ is the total signal, whereas $X \sim \gamma(\alpha, r)$
is the signal of interest, and $Y \sim N(\mu, \sigma^2)$ is the background. 
The \ng\ model was proposed in response to the lack of a proper fit of the
fluorescence 
data with the \en\ model noted by, e.g., \cite{wang2012}, and current applications
of this model rely on a Fast Fourier Transform (FFT) of the convolution. 
The analytic form \eqref{eq:fNG} is expected to provide
a substantial improvement in the analysis of these data both in terms of speed
and accuracy.

Similar applications of the \ng\ distribution can be envisioned for any type of emission that results from the  convolution of signal and background, which is a common experimental
task. For example, background correction in certain biological data
are reviewed by \cite{ritchie2007}, including the \en\ distribution
that is now generalized by the \ng\ distribution. Similar background subtraction
tasks occur routinely in the
physical sciences \citep[see, e.g.,][for applications to astronomy]{ehlert2022, blanton2011}. The \ng\ distribution could therefore provide a suitable and versatile model for
background deconvolution in any application where the two components can be modeled
via these continuous distributions.

\begin{figure}
    \centering
    \includegraphics[width=4in]{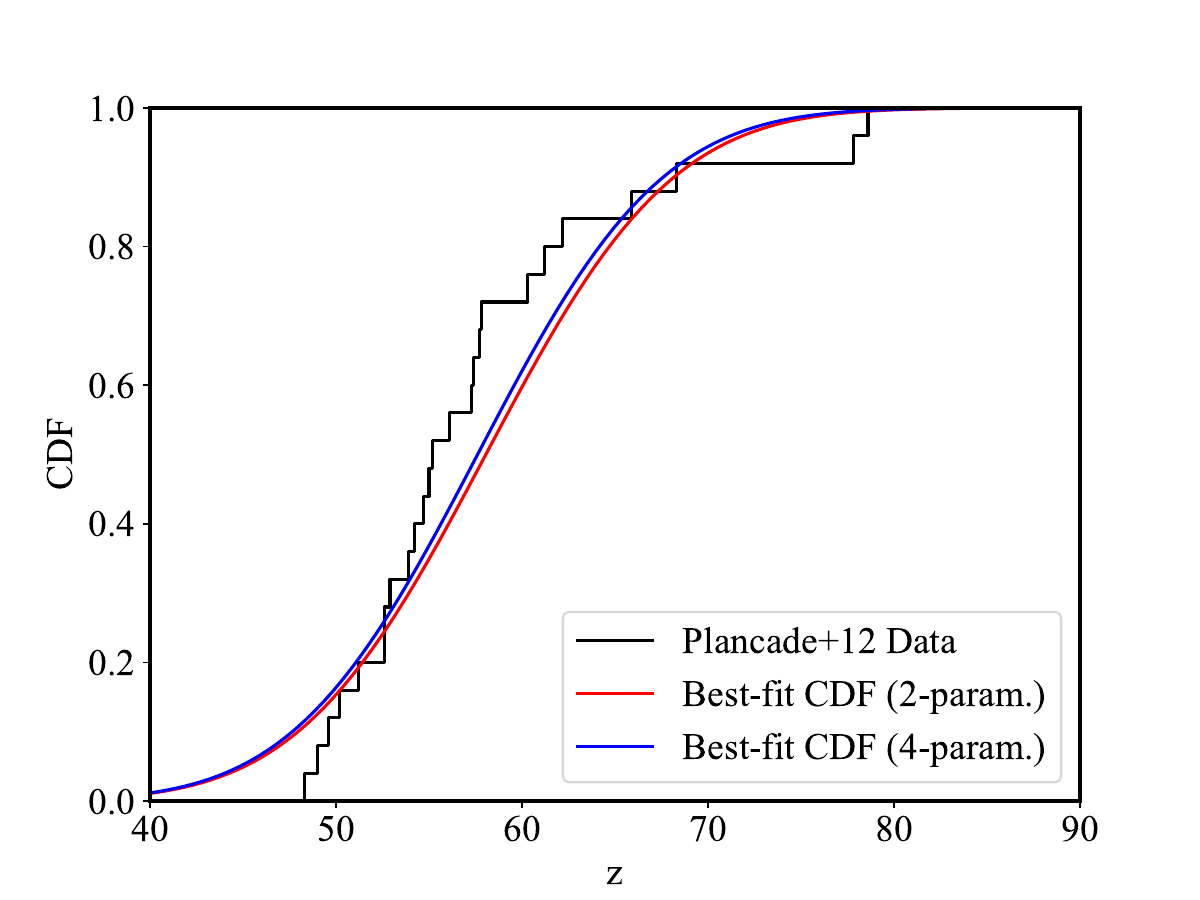}
    \caption{Experimental data from \cite{plancade2012}, and \ml\ regression results with the \ng\ distribution. In
    red is the best--fit CDF using a plug--in estimator for the $(\mu,\sigma)$ parameters, and in blue for the full 4--parameter \ng\ distribution.}
    \label{fig:plancade}
\end{figure}

{
\subsubsection{Application to the \cite{plancade2012} biological data}
The \ml\ methods of this paper are applied to a sub--set of the data analyzed by \cite{plancade2012}. As an example, we 
choose the first $N=25$ measurements from the intensity of a probe from dataset `E1' in that paper, consisting
of a signal that is expected to be the sum of a background term $B$ and a source term $S$. The eCDF of this sub--set of data
for the total signal $T=B+S$ is represented as the black curve, with total probe intensities in the range 48.3--78.6.
The experiment also features an independent measurement of the background $B$, which is expected to be normally distributed.
The background data (not shown in Fig.~\ref{fig:plancade}) have a sample mean and variance of $\hat{\mu}=57.6$ and $\hat{\sigma}=7.8$,
which can be used, following the \cite{plancade2012} notation, as a \emph{plug in} estimator for the two parameters
of the \ng\ distribution, since the background is measured independently from the total signal $T$ (i.e., from different negative probes). These background estimates are consistent with the values reported in \cite{plancade2012}.

From these data we perform two fits. The first is by allowing all 4 parameters of the \ng\ distribution to vary independently,
resulting in the best--fit parameters $(\hat{\alpha},\hat{r},\hat{\mu},\hat{\sigma})=(2.5, 6.7, 54.8, 7.7)$, which result in
the distribution in blue in Fig.~\ref{fig:plancade}. Same as for the numerical simulations described in Sec.~\ref{sec:4paramFit},
this fit results in an observed information matrix  that has a negative determinant, which is due to the parameter identifiability
problem that was discussed in Sec.~\ref{sec:identifiability}, when all parameters are estimated simultaneously. Nonetheless,
convergence of the numerical solution for the score equations is achieved rapidly, and the estimated $\hat{\mu}$ and $\hat{\sigma}$
parameters are in good agreement with the measurements from independent probes. The goodness of fit was measured by the
Kolmogorov--Smirnoff one--sample statistic $D_N=0.21$ \citep{kolmogorov1933}, which corresponds to a null hypothesis probability of $p=0.19$, indicating
good agreement between the eCDF and the best--fit model. 

The second method of regression is obtained by fixing the $(\mu,\sigma)$ parameters at the plug in estimates from the independent probe, and 
proceed with a 2--parameter fit to the $(\alpha,r)$ parameters of the \ng\ distribution that are intended to model
the signal. For this regression we obtain $(\hat{\alpha},\hat{r})=(2.48, 5.65)$, information and covariance matrices

\[
I =  
\begin{bmatrix}
    0.019 & -0.007\\
   -0.007  & 1.758\\
\end{bmatrix},
\hspace{1cm}
\epsilon= 
\begin{bmatrix}
    2.159 & 0.009\\
   0.009  & 0.023\\
\end{bmatrix},
\]

from which we conclude that the two parameters are estimated as $\hat{\alpha}=2.48\pm1.47$ and $\hat{r}=5.65\pm0.15$, with small
positive correlation. This estimate of
the covariance matrix, which was not available in the \ng\ methods of \cite{plancade2012}, is one of the key improvements of
this method of analysis. The Kolmogorov--Smirnov statistic is $D_N=0.23$, i.e., slightly larger than for the 4--parameter fit,
for a null hypothesis probability of $p=0.11$. The KS statistic is indicative of a good agreement between data and model although, as
expected, slightly less so than for the full model. This second method of regression is equivalent to the fits performed by \cite{plancade2012}, whereas the background parameters were estimated by the independent negative probes using the same plug in estimator we used in this analysis.

Some of the challenges of this method of regression are discussed in App.~\ref{sec:appC}. In particular, when values
$z$ from a $Z \sim \ngsymbol ({\alpha},{r},{\mu},{\sigma})$ are large (both positive and negative), the parabolic cylinder functions
and their derivatives become respectively divergent or approach zero (see Fig.~\ref{fig:PCF}). This causes the possibility
of numerical overflow for the functions, and their ratios, which require careful treatment. Possible methods for overcoming these
overflow problems are outlined in App.~\ref{sec:appC}.
}

\subsection{Hypothesis testing of Poisson regression with systematic errors}
\label{sec:applications2}

Systematic errors in the data, broadly defined as additional sources of error due to errors in measurement
or to uncertainties in the underlying model, are a common occurrence in applications
\citep[see, e.g.,][for a review]{glosup1996, vandyk2023}.
An open problem for the use of systematic errors
in data regression is how those errors
affect the goodness--of--fit statistic for integer--valued Poisson data.
The goodness--of--fit statistic for  the regression of Poisson data to a parametric function, also referred to as the Poisson deviance $D_P$,
 is  asymptotically distributed as $\chi^2$ in the large--mean regime
\citep[e.g.,][]{cameron2013,cash1979, bishop1975}, just as it is for normal data (see Sec.~\ref{sec:odchi}).
\cite{bonamente2023} has
proposed a statistical 
model to account for systematic errors
in the \ml\ Poisson regression of certain count data. 
That model results in a goodness--of--fit statistic $Z=X+Y$ that is now distributed as
the sum of the usual $X \sim \chi^2(\nu)$ \DP\ statistic and an independent
$Y \sim N(\mu, \sigma^2)$ statistic
that models the effects of systematic errors, where $\mu$ and $\sigma^2$ can be
estimated from the data.

The \odchi\ distribution  $\odchisymbol(\nu, \mu,\sigma^2)$ 
presented in Sec.~\ref{sec:odchi} is therefore the relevant 
distribution of this Poisson goodness--of--fit statistic, under the null hypothesis of the correctness of the parametric model. Table~\ref{tab:ODC} 
provides the quantiles (i.e., one--sided critical values) of this distribution for selected 
values of the integer number of degrees of freedom $\nu$ and the overdispersion parameter $\sigma^2$. 
These values
are obtained by numerical integration of \eqref{eq:fB}, 
and can be used for the rejection of the
null hypothesis at a given level of significance. Similarly, 
\eqref{eq:fB} can also be used to calculate the $p$--value of the regression, via
$1-p = F_{\odchisymbol}(D_P)$, where $D_P$ is the observed value of the fit statistic and $F_{\odchisymbol}$ denotes the cumulative distribution function.

\begin{figure}
  \centering
  \includegraphics[width=4.2in]{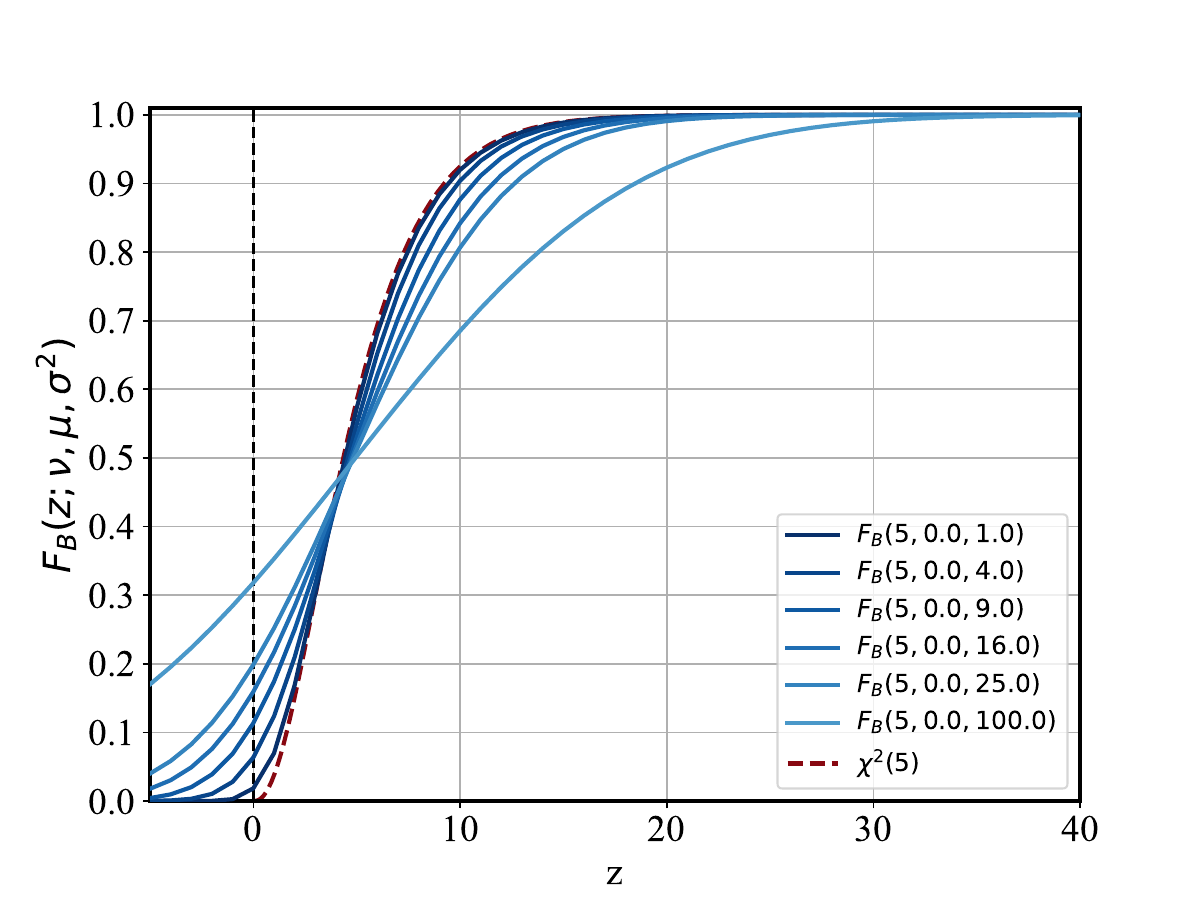}
\caption{CDF of an \odchi\ $B(\nu,\mu,\sigma^2)$ variable for a fiducial value of $\nu$ and $\mu=0$, 
  and for selected values of the overdispersion parameter $\sigma^2$.}
  \label{fig:FBD}
\end{figure}

Figure~\ref{fig:FBD} illustrates an important property of the \odchi\ distribution, namely that 
quantiles of the $B(\nu, \mu=0, \sigma^2)$ distribution for $p>0.5$ are larger than those for the corresponding $\chi^2(\nu)$ distribution, i.e., 
\begin{equation}
    F^{-1}_B(p) > F^{-1}_{\chi^2}(p),\; \text{ for $p>0.5$}.
    \label{eq:P5}
\end{equation}
For the purpose of hypothesis testing, this means that the addition of the normal distribution (as a model for systematic errors) to the chi--square distribution results in larger
critical values, i.e., the presence of systematic
errors makes it more difficult to reject the null hypothesis. Moreover, Property~\ref{pr3}
shows that in the asymptotic limit of a large number of degrees of freedom,
the Poisson deviance is normally distributed, $D_P \sim N(\nu+\mu, 2 \nu +\sigma^2)$,
since the $\chi^2(\nu)$ distribution is approximated by a normal distribution
with the same mean and variance. {
An application of the overdispersed chi--squared distribution for hypothesis testing on real data is reported in \cite{bonamente2024}, to which the interested reader is referred for details.
}

\subsection{Other possible applications}

The gamma distribution for integer values of the 
shape parameter, $r \in \mathbb{N}$, is also known as
the Erlang distribution, after A.~K. Erlang 
who used it to model the distribution
of time between phone calls \citep{erlang1909}. 
The Erlang distribution can be used to model a variety
of counting phenomena that follow from the Poisson
process with fixed rate $\alpha$. Accordingly, the
addition of a normal distribution can be a 
simple way to model overdispersion (and optionally a
shift in the mean, if $\mu \neq 0$) for such processes.

\section{Discussion and conclusions}
\label{sec:discussion}
In this paper, we have first presented key properties of the random variable that
results from the sum of an independent gamma variable with rate parameter $\alpha$ and
shape parameter $r$, and a normal 
random variable with mean $\mu$ and variance $\sigma^2$. The resulting random variable is referred to as the
\ng\ random variable $\ngsymbol(\alpha, r, \mu, \sigma^2)$, and it features a pdf that can be written in
compact form with the aid of parabolic cylinder functions. This new distribution
was first proposed by \cite{plancade2012} for the analysis of certain biological data. 
Two special cases of this distribution are the
\en\ distribution obtained for a fixed shape parameter $r=1$, and the \odchi\ distribution
obtained for a fixed rate parameter $\alpha=\nicefrac{1}{2}$.
A key property of the \ng\ distribution is that the $\mu$ parameter continues to be a simple
location parameter (see Property~\ref{pr1}) that does not otherwise affect the shape of the
distribution, the same as for the normal distribution.

We have then provided analytical expressions for the \ml\ score equations, and the
observed information matrix $I$ of the \ng\ distribution. These results were applied to extensive simulated data sets intended to illustrate the methods of parameter estimation for this class of random variables.
The results provided in this paper suggest that 
of the four parameters of the \ng\ distribution, only three can be identified independently, according
to the Rothenberg rank criterion for $\E(I)$ \citep{rothenberg1971}. This finding in turn
suggests that there may be an alternative reparameterization of the \ng\ model as a function of three
parameters, although such a reparameterization is not obvious to identify. 
When either the rate or the shape parameter of the gamma distribution is held fixed, 
the estimation successfully recovers the other three parameters, providing further indication that
$\rank(I)=3$ for the \ng\ distribution in its most general form. 

Given the broad use of the two
constituting distributions in probability and statistics, the \ng\ distribution
may occur in a wide variety of applications. In particular, the \ng\ distribution was first proposed for
certain biological applications \citep[e.g.][]{plancade2012} as a more general method to fit certain data that were not successfully modeled with the \en\ distribution \citep[e.g.][]{gruskha1972,golubev2010}. It is hoped that the present
investigation, with the derivation of analytic expressions for the \ml\ estimation, will benefit such
applications.

The \odchi\ distribution, on the
other hand, is the parent model for the goodness--of--fit of certain Poisson data that
feature systematic errors \citep[e.g.][]{bonamente2023}. In that class of applications, 
the main difference with the positively--supported $\chi^2$ distribution
 is a tail of negative values that becomes more prominent as $\sigma$ increases, for a fixed value of $\mu$,
 and it is a consequence of the convolution with the real--valued normal distribution that models
 the presence of systematic errors.
The paper has also provided critical
values for the $B(\nu, \mu=0, \sigma^2)$ for representative values of the $\nu$ and $\sigma^2$
parameters that can be used for hypothesis testing.~\footnote{\python\ codes to reproduce the results presented in this
  paper, including the new \texttt{NG} and \texttt{OChi2} classes of continuous distributions, and all
  associated functions, are available
  for use on the first author's \github\ page at \texttt{https://github.com/bonamem}.} Thanks to the translation
  property of the \ng\ distribution, these quantiles can be used also for applications with $\mu \neq 0$.

\section*{Declarations}
The authors declare no conflicts of interest or competing interests for this manuscript.

\begin{appendices}

\section{Distributions and functions of interest}
\label{sec:appA}

The probability distribution function (pdf) of a gamma random variable $ X \sim \gamma(\alpha, r)$
is defined as
\begin{equation}
    f_{\gamma}(x) = \dfrac{\alpha (\alpha x)^{r-1}}{\Gamma(r)} e^{-\alpha x}
\end{equation}
with $\alpha, r$ positive real numbers, and $x \geq 0$. The parameter $\alpha$ is
referred to as the rate parameter (and $1/\alpha$ as the scale parameter), and $r$ is the shape parameter.

The pdf of an exponential random variable $X \sim \mathrm{Exp}(\alpha) $ has a probability distribution 
\begin{equation}
    f_E(x) = \alpha e^{-\alpha x}
\end{equation}
where $\alpha$ is known as the rate parameter, and $1/\alpha$ as the scale parameter.

The pdf of a chi--squared random variable $X \sim \chi^2(\nu)$ is 
\begin{equation}
    f_{\chi^2}(x) = \left(\dfrac{1}{2}\right)^{\nu/2} \dfrac{1}{\Gamma(\nu/2)} e^{-x/2} x^{\nu/2-1}
\end{equation}
with $\nu \in \mathbb{R}^+$.  In many statistical applications, $\nu \in \mathbb{N}$ signifies the number of degrees of freedom of the distribution.

The error function is defined as
\begin{equation}
    \erf(x) = \dfrac{2}{\sqrt{\pi}} \int_{0}^{\infty} e^{-t^2} dt.
\end{equation}
Let $F(x)$ be the cumulative distribution function of a standard normal variable $X \sim N(0,1)$. Then it is true that
\begin{equation}
    F(\sqrt{2}\, x) = \dfrac{1}{2} \left (1+ \erf(x) \right)
\end{equation}
and
\begin{equation}
    F(- \sqrt{2}\, x) = \dfrac{1}{2} \left (1 - \erf(x)\right).
\end{equation}
The latter equation can be used to show that the pdf of the \en\ distribution as
given by \eqref{eq:fEG} is equivalent to Eq.~3 of \cite{xie2009}.

The following recursion formulas apply to the parabolic cylinder functions \citep[see 9.247,][]{gradshteyn2007}:
\begin{equation}
    \begin{cases}
        D_{p+1}(z) - z D_p(z) + pD_{p-1}(z) =0\\[10pt]
        \dfrac{d}{dz} D_p(z) = \dfrac{1}{2} z D_p(z) - D_{p+1}(z)\\[10pt]
         \dfrac{d}{dz} D_p(z) = -\dfrac{1}{2} z D_p(z) + p D_{p-1}(z)
    \end{cases}
    \label{eq:DpRecursion}
\end{equation}
In particular, given the restriction that the index of the parabolic cylinder function 
is negative, the last equation of \eqref{eq:DpRecursion}  applies when $p-1<0$. With our use
of the parameter $r=-p$, this restriction translates to $r >-1$, which is always satisfied in our application. Therefore
the equation of choice to express the derivative of the parabolic cylinder functions
with respect to its argument, via the parabolic functions themselves, becomes
\begin{equation}
    \dfrac{d}{d \zeta} D_{-r}(\zeta) = -\dfrac{\zeta}{2} D_{-r}(\zeta) 
    - r D_{-r-1}(\zeta).
    \label{eq:DpRecursion2}
\end{equation}

\section{Equations for maximum--likelihood estimation with the \ng\ distribution}
\label{sec:appB}
Consider $N$ iid measurements $z_i \sim \ngsymbol(\theta)$, with $\theta=(\alpha, r, \mu, \sigma)$
the four parameters of the \ng\ distribution. The log--likelihood is
\begin{equation}
    \mathcal{L}= \sum_{i=1}^N \ln f_{\ngsymbol}(z_i/\theta)
    \label{eq:L}
\end{equation}
with 
\begin{equation}
    \ln f_{\ngsymbol}(z_i/\theta) = \ln \left(\dfrac{\alpha^r \sigma^{r-1}}{\sqrt{2 \pi}} \right)
    + \ln D_{-r}(\zeta) + \ln E(z),
\end{equation}
where $\zeta(z)$ is given by \eqref{eq:zeta} and $\ln E(z)$ by \eqref{eq:E}.

\subsection{The score function}
The maximum likelihood estimates $\hat{\theta}$ are given by setting the score functions to zero, $d \mathcal{L}/d \theta|_{\hat{\theta}} = 0$, and then solving. We will consider separately the
score functions for the parameters $\alpha, \mu, \sigma$, and then the one for the parameter $r$ which occurs
as the index of the parabolic cylinder function.

(A) For $\theta=(\alpha, \mu, \sigma)$, the derivatives of the log--likelihood are obtained from derivatives of elementary functions,
\begin{equation}
    \begin{cases}
        \dfrac{d \zeta}{d \theta} = \left(\sigma, \dfrac{1}{\sigma}, \alpha - \dfrac{\mu -z}{\sigma^2}\right)\\[10pt]
        \dfrac{d \ln E(z)}{d \theta} = \left( 0, \dfrac{z-\mu}{\sigma^2},  \dfrac{(z-\mu)^2}{\sigma^3} 
       \right) + \dfrac{\zeta}{2} \left(\dfrac{d \zeta}{d \theta} \right),
    \end{cases}
\end{equation}
and from the derivative of the parabolic cylinder function with respect to its argument, according to the bottom equation of \eqref{eq:DpRecursion}, which is always satisfied in this application:
\begin{equation}
    D_{-r}^'(\zeta) = 
          -\dfrac{1}{2} \zeta D_{-r}(\zeta) -r D_{-r-1}(\zeta), \text{ if $r\geq -1$}.
    \label{eq:Dpprime}
\end{equation}

Therefore, for the parameters $\theta=(\alpha, \mu, \sigma)$, the three
score functions for the $i$--th datum are summarized as
\begin{equation}
    \left. \dfrac{d \ln f_{\ngsymbol}(z)}{d \theta} \right|_{z_i} = 
    \left( \dfrac{r}{\alpha}, 0, \dfrac{r-1}{\sigma} \right) +
    \dfrac{D_{-r}^'(\zeta_i)}{D_{-r}(\zeta_i)} \, \left. \dfrac{d \zeta}{d \theta} \right|_{z_i} +  
    \left. \dfrac{d \ln E(z)}{d \theta}\right|_{z_i}. 
    \label{eq:d1}
\end{equation}

(B) For $\theta=r$, a derivative of $D_{-r}(\zeta)$ with respect to $r$ is required. With
\begin{equation}
  D_{-r}(z) = \dfrac{e^{-z^2/4}}{\Gamma(r)} \int_0^{\infty} e^{-xz -x^2/2} x^{r-1} dx 
  \equiv \dfrac{e^{-z^2/4}}{\Gamma(r)} A(z,r),
  \label{eq:Dr}
\end{equation}
the problem becomes that of finding the derivatives of $A(z,r)$ and $\Gamma(r)$ with respect to r.
First, recall that
\begin{equation*}
    \dfrac{d \Gamma(r)}{d r} = \psi(r) \Gamma(r)
\end{equation*}
where $\psi(r)$ is the digamma function. For $A(z, r)$, we may use the Leibnitz rule to obtain
\begin{equation} 
\dfrac{\partial A(z, r)}{\partial r} = \int_0^{\infty} e^{-xz -x^2/2} x^{r-1} \ln x\,  dx  \equiv A_r(z,r)
\label{eq:Ar}
\end{equation}
which does not appear to have a simpler analytic form. Therefore
\begin{equation}
\dfrac{\partial}{\partial r} \ln D_{-r}(\zeta) = \left. \dfrac{A_r(\zeta, r)}{A(\zeta,r)} \right|_{z_i} - \psi(r),
\label{eq:DrAAr}
\end{equation}
and the score function for $\theta=r$ is given by
\begin{equation}
   \left. \dfrac{\partial \ln f_{\ngsymbol}(z)}{\partial r} \right|_{z_i} = 
   (\ln \alpha+\ln \sigma) + \left. \dfrac{A_r(\zeta, r)}{A(\zeta,r)} \right|_{z_i} - \psi(r).
   \label{eq:d2}
\end{equation}
Note how the factorization of \eqref{eq:Dr} is convenient for the logarithmic derivative.

Accordingly, the score equations for $\hat{\theta}_j$, for $j=1,\dots, 4$, are
\begin{equation}
    \dfrac{\partial \mathcal{L}}{\partial \theta_j}=\sum_{i=1}^N \left. \dfrac{\partial \ln f_{\ngsymbol}(z)}{\partial \theta_j} \right|_{z_i, \hat{\theta}} = 0,
    \label{eq:score}
\end{equation}
with derivatives given by \eqref{eq:d1} and \eqref{eq:d2}. Eq.~\ref{eq:score}
is a system of four nonlinear equations that requires numerical solution for $\hat{\theta} = 
(\hat{\alpha}, \hat{r}, \hat{\mu}, \hat{\sigma})$. When evaluating the sum of the scores over $N$ iid measurements, it is convenient to define the following sums:

\begin{equation}
    \begin{cases}
    Z_1 = \dfrac{1}{N}\sum\limits_{i=1}^N (\mu - z_i) = (\mu - \overline{z}),\; Z_2=  \dfrac{1}{N}\sum\limits_{i=1}^N (\mu - z_i)^2;\\[10pt]   
    S = \dfrac{1}{N}\sum\limits_{i=1}^N \DpDi,\; S_{\mu} = \dfrac{1}{N}\sum\limits_{i=1}^N \DpDi (\mu-z_i); \\[15pt]
        T = \dfrac{1}{N} \sum\limits_{i=1}^N \ArAi. \\[10pt] 
    \end{cases}
    \label{eq:SumsScore}
\end{equation}

\subsection{The information matrix}

The observed information matrix is the symmetric $4 \times 4$ matrix
\[
I = -\dfrac{\partial^2 \mathcal{L}}{\partial \theta \partial \theta^T} = -\dfrac{\partial}{\partial \theta}
\left( \dfrac{\partial \mathcal{L}}{\partial \theta^T} \right) = -\sum_{i=1}^N \dfrac{\partial^2}{\partial \theta \partial \theta^T} \ln f_{\ngsymbol}(z_i/\theta)
\]
with $\theta=(\alpha, r, \mu, \sigma)$ and $T$ denoting the transpose of a vector, and therefore it is the sum over the $N$ measurements of the second--order partial derivatives of the logarithm of the \ng\ density. 

The second--order derivatives can be calculated starting from \eqref{eq:d1} and \eqref{eq:d2}. The derivatives with respect to $r$ (the index of the parabolic cylinder functions) and with respect 
to the other parameters that appear in the argument $\zeta=\zeta(\alpha, \mu, \sigma)$ are calculated
separately, given that the recursion formulas \eqref{eq:Dpprime} only apply to
derivatives
with respect to the argument $\zeta$. The second--order derivatives require the following functions:

\begin{equation}
    \dfrac{d}{d \zeta} \left[ \dfrac{D_{-r}'(\zeta)}{D_{-r}(\zeta)}\right] = \dfrac{d}{d \zeta} \left[ \dfrac{d}{d \zeta} \ln D_{-r}(\zeta) \right] = \dfrac{D_{-r}''(\zeta) D_{-r}(\zeta) - D_{-r}'(\zeta)^2}{D_{-r}(\zeta)^2}
\end{equation}
where the recursion equations \eqref{eq:Dpprime} can be used to evaluate the first--order derivatives $D_{-r}'(\zeta)$ as a function of the parabolic cylinder functions, see \eqref{eq:Dpprime}. Moreover,
\begin{equation}
    D_{-r}''(\zeta) \equiv \dfrac{\partial}{\partial \zeta} D_{-r}'(\zeta)=
       -\dfrac{1}{2} D_{-r}(\zeta) -\dfrac{\zeta}{2}D_{-r}'(\zeta) - r D_{-r-1}'(\zeta)  \; \text{ for $r \geq -1$}
    \label{eq:Dpsecond}
\end{equation}
where the recursion equations \eqref{eq:Dpprime} were used. Given that the derivatives of the parabolic cylinder functions are readily available, it is not necessary to use the recursion equations again in \eqref{eq:Dpsecond} in order to eliminate the derivatives. 

For the derivatives with respect to $r$, the following results are needed. First,
\begin{equation}
    \dfrac{\partial}{\partial r} \left[ \dfrac{D_{-r}'(\zeta)}{D_{-r}(\zeta)} \right] = 
    \dfrac{\dfrac{d}{d r}D_{-r}'(\zeta) \cdot D_{-r}(\zeta)  - \dfrac{d}{dr} D_{-r}(\zeta) \cdot  D_{-r}'(\zeta) }{D_{-r}(\zeta)^2 }
\end{equation}
where the derivatives with respect to $r$ are given by
\[
\begin{cases}
    \dfrac{d}{dr} D_{-r}(\zeta) = D_{-r}(\zeta) \left( \dfrac{A_r(\zeta,r)}{A(\zeta,r)} - \psi(r)\right)\\[10pt]
    \dfrac{d}{dr} D_{-r}'(\zeta) = 
        -\dfrac{1}{2} \zeta  \dfrac{d}{dr} D_{-r}(\zeta) - D_{-r-1}(\zeta) - r \dfrac{d}{dr} D_{-r-1}(\zeta)\; \text{ for $r\geq -1$}
\end{cases}
\]
and therefore they can all be expressed as functions of the parabolic cylinder functions
and the function $A_r(\zeta,r)$ that was defined in \eqref{eq:Ar}.

The derivatives with respect to $r$ of the two functions of $r$ in \eqref{eq:d2} can be evaluated as
\begin{equation}
\dfrac{d}{dr} \left( \dfrac{A_r(\zeta,r)}{A(\zeta,r)} \right) = \dfrac{A_{rr}(\zeta,r) A(\zeta,r) - A_r(\zeta,r)^2}{A(\zeta,r)^2}
\end{equation}
which requires the following additional integral:
\begin{equation}
    \dfrac{d}{dr} A_r(\zeta,r) =  \int_0^{\infty} e^{-xz -x^2/2} x^{r-1} (\ln x)^2\,  dx  \equiv A_{rr}(z,r).
\label{eq:Arr}
\end{equation}

The cross--derivative of $A_r/A$ with respect to $\zeta$ can be related to the cross--derivative of $\partial/
\partial \zeta \ln D_{-r} (\zeta)$ with the use of \eqref{eq:DrAAr} and a change in the order 
of the derivatives:
\[
\dfrac{\partial}{\partial \zeta} \left(\frac{A_r}{A}\right) = \dfrac{\partial}{\partial \zeta} \left(
\dfrac{\partial}{\partial r} \ln D_{-r}(\zeta)\right) =   \dfrac{\partial}{\partial r} \left(
\dfrac{\partial}{\partial \zeta} \ln D_{-r}(\zeta)\right).
\]

Finally, the derivative of the digamma function is a polygamma function of order 1, $\psi_1(r)$,
\[
\dfrac{d}{dr} \psi(r) = \psi'(r) = \psi_1(r).
\]



The second--derivatives matrix $I_{\star}=(I_{1\,\star}^T,I_{2\,\star}^T,I_{3\,\star}^T,I_{4\,\star}^T)$ for one observation of the \ng\ distribution is as follows:
\begin{equation} 
\begin{aligned}
    I_{1\,\star}= & \dfrac{d}{d \theta } \left(\dfrac{\partial}{\partial \alpha} \ln f_{\ngsymbol}(z_i) \right) =
 \left( - \dfrac{r}{\alpha^2} + \dfrac{d}{d \alpha}\DpD \sigma
    + \dfrac{\sigma^2}{2},\dfrac{1}{\alpha} + \sigma \dfrac{\partial}{\partial r} \DpD, \right. \\
    & \left. \dfrac{d}{d \mu} \DpD \sigma+ \dfrac{1}{2},
    \dfrac{d}{d \sigma} \DpD \sigma  + \DpD + \alpha \sigma \right)
\end{aligned}
\label{eq:I1}
\end{equation}

    \begin{equation}
    \begin{aligned}
    I_{2\,\star}= &  \dfrac{d}{d \theta } \left(\dfrac{\partial}{\partial r} \ln f_{\ngsymbol}(z_i) \right) =  \left( \dfrac{1}{\alpha} + \dfrac{d}{d \alpha} \left[ \ArA \right], \dfrac{d}{dr} \left[ \ArA - \psi(r) \right]
     , \right. \\
    & \left. \dfrac{d}{d \mu}  \left[ \ArA \right], 
    \dfrac{1}{\sigma} \dfrac{d}{d \sigma} \left[ \ArA \right] \right)
    \end{aligned}
    \label{eq:I2}
    \end{equation}

\begin{equation} 
\begin{aligned}
    I_{3\,\star}= & \dfrac{d}{d \theta } \left(\dfrac{\partial}{\partial \mu} \ln f_{\ngsymbol}(z_i) \right) = \\
    & \left(  \dfrac{d}{d \alpha} \DpD \dfrac{1}{\sigma} + \dfrac{1}{2}, 
    \dfrac{d}{dr} \DpD \dfrac{1}{\sigma},
    \dfrac{d}{d \mu} \DpD \dfrac{1}{\sigma} - \dfrac{1}{2 \sigma^2}, \right . \\
    & \left .\dfrac{d}{d \sigma} \DpD \dfrac{1}{\sigma} - \dfrac{1}{\sigma^2} \DpD  
    - \dfrac{z-\mu}{\sigma^3} - \dfrac{\alpha}{\sigma} \right)
\end{aligned}
\label{eq:I3}
\end{equation}

\begin{equation} 
\begin{aligned}
    I_{4\,\star}= & \dfrac{d}{d \theta } \left(\dfrac{\partial}{\partial \sigma} \ln f_{\ngsymbol}(z_i) \right) =  \left( \dfrac{d}{d \alpha} \DpD \dzdsigma + \DpD + \alpha \sigma, \right. \\
&  \dfrac{1}{\sigma} + \dfrac{d}{dr} \DpD \dzdsigma,
\dfrac{d}{d \mu} \DpD \dzdsigma \\ &  - \dfrac{1}{\sigma^2} \DpD 
+ \dfrac{\mu - z}{\sigma^3},\\
 & - \dfrac{r-1}{\sigma^2} + \dfrac{d}{d \sigma} \DpD \dzdsigma + \DpD \dfrac{2(\mu-z)}{\sigma^3} 
+ \dfrac{\alpha}{2} \\
 & \left. - \dfrac{3}{2} \dfrac{(z-\mu)^2}{\sigma^4}  \right)
\end{aligned}
\label{eq:I4}
\end{equation}

When evaluating the information matrix for $N$ iid measurements, as in \eqref{eq:I1text}, all the functions in
\eqref{eq:I1}--\eqref{eq:I4} must be evaluated at the $i$--th measurement
$\zeta(z_i)$ according to \eqref{eq:zeta}.

The following sum over the $N$ data points must be evaluated to calculate the information matrix:
\begin{equation}
\begin{cases}
    S_{\zeta} =  \dfrac{1}{N}\sum\limits_{i=1}^N \dzdDi = \dfrac{\partial S}{\partial \zeta}, \\[10pt]
    S_r = \dfrac{1}{N}\sum\limits_{i=1}^N \left[ \dfrac{\partial}{\partial r} \DpDi \right] = \dfrac{\partial S}{\partial r}, \\[10pt]
    S_{r,\mu}=  \dfrac{1}{N}\sum\limits_{i=1}^N \left[ \dfrac{\partial}{\partial r} \DpDi (\mu-z_i)\right],\\[10pt]
    T_r  = \dfrac{1}{N} \sum\limits_{i=1}^N \left[ \dfrac{\partial}{\partial r} \ArAi \right] = \dfrac{\partial T}{\partial r},\\[10pt]
    S_{\zeta,\mu} =  \dfrac{1}{N}\sum\limits_{i=1}^N \dzdDi (\mu-z_i),\\[10pt]
    S_{\zeta,\mu^2} =  \dfrac{1}{N}\sum\limits_{i=1}^N \dzdDi (\mu-z_i)^2.\\[10pt]
\end{cases} 
\label{eq:SumsI}
\end{equation}

\section{Numerical methods for \ml\ estimation}
\label{sec:appC}

\subsection{Iterative solution of the score equations}

Numerical solution of the score equations \eqref{eq:scoreShort} was performed with the
\texttt{least_squares} function available from \texttt{scipy} in \texttt{python}, which
seeks a solution of the score equations.
Although the Jacobian of the system of equations
is available as the second--order derivatives of the log likelihood in \eqref{eq:I1text},
we opted for a numerical estimation via the \texttt{2-point} optional parameter.

{We find that the starting point of the numerical solution for the parameters $(\alpha, r, \mu, \sigma)$
using \texttt{least_squares} is typically not a crucial step in the analysis. For the simulations, we started at the parent values, and 
also tested that starting points with large variations ($\geq 100$\%) from the parent values also quickly converge to the solution. For the real data example of Sec.~\ref{sec:applications1}, the numerical solution of the score equations started at nominal values of $\alpha=1$ and $r=2$, and convergence to
the best--fit values was also rapidly achieved in few steps.} 

{\subsection{Evaluation of parabolic cylinder functions and related functions}

The main numerical challenges for this analytic method of \ml\ estimation are associated with
the large values of the parabolic cylinder functions (see Fig.~\ref{fig:PCF}). In this section we
detail the numerical methods used in this paper, and outline some of the challenges and possible
ways to overcome them in applications.
}

\subsubsection{The ratio $D'_{-r}(\zeta)/D_{-r}(\zeta)$ and its derivative with respect to $\zeta$}
\label{sec:appC2}
A numerical challenge to finding the solution of the score
equations is the large value of the parabolic cylinder functions $D_{-r}(\zeta)$ and its derivative
with respect to the argument $\zeta$, i.e. $D'_{-r}(\zeta)$, {when $z$ is large and positive; and 
the small values for these functions when $z$ is large and negative, since $D_{-r}(\zeta)$ appears at the denominator of
certain functions needed for estimation, e.g., \eqref{eq:SumsScore} and \eqref{eq:SumsI}.}
{Both the parabolic cylinder function and its derivatives with respect to $\zeta$ are} 
 available in the same package via the \texttt{pbdv} function. 
As illustrated in Fig.~\ref{fig:PCF}, the parabolic cylinder 
functions diverge as $z$ increases, and therefore as $\zeta$ decreases according to
\eqref{eq:zeta}; similar asymptotic divergence occurs for the
derivative, which is related to the function itself via the recursion relations
\eqref{eq:DpRecursion}. Accordingly,  datasets with large values of the $Z$ variable{, both positive and negative,} need to
be handled with care due to possible overflow problems. 

Fortunately, it is possible to overcome these problems for most applications. 
We suggest two possible avenues.\\
(a) According to the translation property of the \ng\ distribution (Property~\ref{pr1}), the data can be shifted to lower values
by a fixed constant. This constant is then added to the $\mu$ parameter estimate. \\
(b) The scores and the information matrix depend only on logarithmic derivatives, i.e., the ratios of the
derivatives to the parabolic cylinder functions themselves, see the sum terms $S$ and $S_{\zeta}$ in \eqref{eq:SumsScore} and \eqref{eq:SumsI}. Both ratios $D'_{-r}(\zeta)/D_{-r}(\zeta)$
and $d/d \zeta (D'_{-r}(\zeta)/D_{-r}(\zeta))$ have values that are close to unity, in a large range of the variable $\zeta$. In particular, we observe that, {for all $r$,}
\[
\lim\limits_{\zeta \to {\color{blue}\pm} \infty} \; \dfrac{d}{d \zeta} \left(
\dfrac{D'_{-r}(\zeta)}{D_{-r}(\zeta)} \right) = {\mp} \dfrac{1}{2},
\]
which can be used for the sums $S_{\zeta}$, $S_{\zeta,\mu}$ and $S_{\zeta, \mu^2}$ in the asymptotic limit of a large argument. 
There are also a number of asymptotic expansions
of the parabolic cylinder function \citep[see Sec. 9.246 of][]{gradshteyn2007}
that can be used to approximate the sums $S$ and $S_{\mu}$ for large values of their 
argument.
It is therefore possible to use asymptotic expressions for the ratios in the sums
\eqref{eq:SumsScore} and \eqref{eq:SumsI} to bypass the direct evaluation of the parabolic cylinder functions, and to avoid overflow problems. Those approximations, {which in general may depend of the value of $r$,} are not discussed in this paper.

\subsubsection{The ratio $A_r(\zeta,r)/A(\zeta,r)$ and its derivative with respect to~$r$}

The derivatives of the parabolic cylinder functions with respect to the index $r$
lead to the integrals $A_r$ and $A_{rr}$ that are evaluated according to the Leibnitz rule;
see  \eqref{eq:Ar} and \eqref{eq:Arr}.
Those integrations can be performed via standard numerical integration methods,
and they {become large} 
for large negative values of the $Z$ variable, and accordingly for
large positive values of the $\zeta$ argument. {This is a numerical challenge
that may affect datasets with large negative values of the \ng\ variable.}
Fortunately, the ratio $A_r(\zeta,r)/A(\zeta,r)$ remains a small number in the vicinity of zero for a large range of the $\zeta$ and $r$ parameters. The same applies to the ratio 
$d/dr (A_r(\zeta,r)/A(\zeta,r))$ which appears in $T_r$. 
{Therefore, for applications in this regime of the \ng\ variable, it may be convenient to seek numerical approximations for the asymptotic values
of these ratios.}
{Approximations for the $A$, $A_r$ and $A_{rr}$ integrals are not discussed in this paper.}

\end{appendices}

\bibliography{max,maxODChi}
\end{document}